\documentclass{iopjournal}

\usepackage[T1]{fontenc}
\usepackage[utf8]{inputenc} 

\usepackage{amsmath,amssymb,amsfonts}

\usepackage{tikz}
\usetikzlibrary{
  arrows.meta,
  positioning,
  calc,
  fit,
  backgrounds
}

\usepackage[percent]{overpic}

\usepackage[protrusion=true,expansion=true,final]{microtype}

\hypersetup{
  colorlinks=true,
  linkcolor=blue,
  citecolor=blue,
  urlcolor=blue
}

\newcommand{\IOPjournalname}{Reports on Progress in Physics}

\makeatletter
\renewcommand{\articletype}[1]{%
{\vspace*{-8mm}\noindent \Large \sf \IOPjournalname}

\vspace*{8mm} \noindent\reversemarginpar
\marginpar{%
\vspace{-3mm}
{\color{gray}\hrule} \ \\ Crossmark\\
{\color{gray}\hrule} \ \\
\tiny {\sf RECEIVED} {\small \\ dd Month yyyy}\\ \\
{\sf REVISED} {\small \\ dd Month yyyy}%
}%
{\scriptsize \sf{\bfseries \MakeUppercase{#1}}}%
}
\makeatother

\begin{document}

\articletype{Report on Progress}

\title{What Makes Three-Dimensional Quantum Spin Liquids Possible?}

\author{Yasir Iqbal$^{1,*}$ and Ronny Thomale$^{2,1}$}

\affil{$^1$Department of Physics, Indian Institute of Technology Madras, Chennai 600036, India}

\affil{$^2$Institut f\"ur Theoretische Physik und Astrophysik, and W\"urzburg-Dresden Cluster of Excellence ctd.qmat, Julius-Maximilians-Universit\"at W\"urzburg, Am Hubland, Campus S\"ud, 97074 W\"urzburg, Germany}

\affil{$^*$Author to whom any correspondence should be addressed.}

\email{yiqbal@physics.iitm.ac.in}


\begin{abstract}
Quantum spin liquids are often introduced through a low-dimensional intuition:
weak coordination and strong zero-point motion frustrate conventional magnetic
order. This view has shaped much of the field, but it misses another route to
quantum disorder. In many frustrated magnets, the more natural starting point is
not a fluctuating version of an ordered state, but a locally constrained manifold
of low-energy configurations. Within such a manifold, quantum dynamics can
generate an emergent gauge theory, with fractionalized excitations and collective
modes absent in ordinary magnets. Here we ask why such phases can be stable in three spatial dimensions. We argue that three-dimensional quantum spin liquids need not be regarded as fragile exceptions to the tendency of 3D magnets to order. They can arise when local constraints suppress premature order selection, coherent tunneling processes
connect the constrained manifold, and the topology of gauge defects permits a
deconfined regime. For compact gauge theories, three dimensions can even reverse
the usual dimensional intuition: regimes that are unstable or fine-tuned in
lower dimensions may become stable over finite regions of parameter space. We organize the discussion around constraint-driven Coulomb phases, weak harmonic order-by-disorder selection, and the role of spin--orbit or multipolar interactions in generating the required microscopic dynamics. We close with experimental diagnostics, candidate materials, and open theoretical questions for three-dimensional quantum spin liquids.
\end{abstract}

\tableofcontents

\section{Three dimensions and the stability of quantum spin liquids}


Quantum spin liquids (QSLs) are often introduced as phases stabilized by unusually strong quantum fluctuations, and are therefore heuristically associated with low dimensionality. This view captures an important part of the story, but it is incomplete, because local constraints can reorganize the low-energy Hilbert space before conventional ordering tendencies become decisive. We argue here that three dimensions do not merely tolerate QSLs—they can \emph{stabilize} them by changing the infrared topology of the emergent gauge sector. When dominant interactions impose local constraints (simplex, ice-rule, or Gauss-law type), the system does not fluctuate about a unique classical configuration; instead it explores an extensively degenerate manifold that is most naturally described by gauge variables~\cite{Balents2010Nature,SavaryBalents2016RoPP}. Quantum dynamics that act \emph{within} this constrained manifold (transverse exchange, multipolar tunneling, or multi-spin ring processes) can then recast the low-energy problem as a compact lattice gauge theory.

In effective low-energy descriptions we will distinguish a ``pure'' gauge theory (only gauge-field degrees of freedom retained) from a gauge theory coupled to dynamical matter fields. In microscopic lattice spin systems, additional excitations are always present at some scale: spinons carry gauge charge, magnetic monopoles occur in compact U(1) descriptions, and visons represent gauge-flux excitations in $\mathbb{Z}_2$ gauge theories. In controlled regimes the charged matter fields may be parametrically gapped and integrated out, leaving an approximately pure gauge theory governing the infrared gauge sector~\cite{Fradkin-2013}. When matter remains low-lying, its symmetry and dynamics can qualitatively modify confinement and finite-$T$ behavior, so we will indicate explicitly when a discussion assumes gapped matter versus active matter fields. This broader viewpoint, in which the structure of the low-energy sector and its defects governs the resulting topological order and emergent gauge fields, has also been emphasized in more general settings~\cite{Wen-1990,Wen-2002,Sachdev-2019}.

The distinction between two and three dimensions is sharpest for compact U(1) gauge
structure without gapless matter. In two spatial dimensions, such theories admit monopole-instanton events that generically proliferate and confine the would-be deconfined phase~\cite{Polyakov-1977}. In three spatial dimensions, by contrast, compact U(1) gauge theory admits a stable weak-coupling Coulomb phase, characterized by a gapless emergent photon and gapped magnetic monopoles \cite{HermeleFisherBalents2004PRB}. Thus the question ``what makes three-dimensional QSLs possible?'' is not answered by dimensionality alone, but by the combination
of a constraint-generated gauge structure and an infrared gauge theory whose defects do not immediately confine it.

One can trace a much earlier prehistory of quantum-disordered
magnetism to Landau’s circle: in 1941, Pomeranchuk proposed
that a non-antiferromagnetic insulating magnet might support
neutral fermionic excitations, an idea later recalled and
reinterpreted by Dzyaloshinskii in the context of strongly
correlated Mott physics \cite{Pomeranchuk-1941,Dzyaloshinskii-1989}.
One may also note an early, largely forgotten precursor in the
discussion of singlet-based magnetism: Rudnitskii (1942) criticized
the conventional checkerboard picture of antiferromagnetism and argued
instead that the ground state could correspond to a ``molecular''
electronic structure with nonlocalized valence bonds and antiparallel
spins~\cite{Rudnitskii-1942}. Although this remained far removed from
the modern language of spin liquids, it is striking in retrospect as an
early anticipation of the idea that an antiferromagnet might admit a
singlet ground state not captured by conventional magnetic order.
What in Rudnitskii still appeared only as an isolated and largely
pre-conceptual singlet-based alternative was elevated by Anderson into
a general many-body principle: the proposal that a quantum magnet could
realize a resonating valence bond (RVB) ground state, not merely as a
bonded alternative to N\'eel order, but as a new kind of
insulator~\cite{Anderson-1973}. This early RVB intuition was quickly sharpened in concrete model studies. In particular, S\"ut\H{o} and Fazekas compared the square- and triangular-lattice $S=\tfrac12$ Heisenberg antiferromagnets and argued that, unlike the square lattice, the triangular case could plausibly support a singlet, spin-liquid-like ground state rather than conventional N\'eel order~\cite{Suto-1977}. Very early on, the RVB framework was also recognized as carrying nontrivial topological structure and fractionalized
excitations, including neutral spin-$\tfrac12$ fermions and
charge-$\pm e$ spinless bosons, in work by Kivelson,
Rokhsar, and Sethna on the topology of the RVB
state~\cite{Kivelson-1987} as well as, inspired by an idea first put forward by Dung-Hai Lee, by Kalmeyer and Laughlin on chiral spin liquids which are fractional quantum Hall states transposed into a Spin-$S$ Hilbert space~\cite{PhysRevLett.59.2095,PhysRevLett.102.207203}. This line of thought was later systematized in the modern language of topological order and quantum order, which clarified how spin liquids may evade characterization in terms of conventional symmetry breaking alone~\cite{Wen-1990,Wen-1991,Wen-2002}. The revival of RVB ideas
in the context of high-temperature superconductivity
reframed quantum spin liquids as parent states of doped
Mott insulators~\cite{Anderson-1987}. Early large-$N$
analyses of frustrated quantum antiferromagnets further
sharpened this RVB/valence-bond viewpoint and made
clear that frustration could lead not only to ordered
states selected by fluctuations, but also to genuinely
quantum-disordered phases and valence-bond order
\cite{Read-1989,Read-1991,Sachdev-1992}.

\begin{figure}[t]
    \centering
    \includegraphics[width=0.92\textwidth]{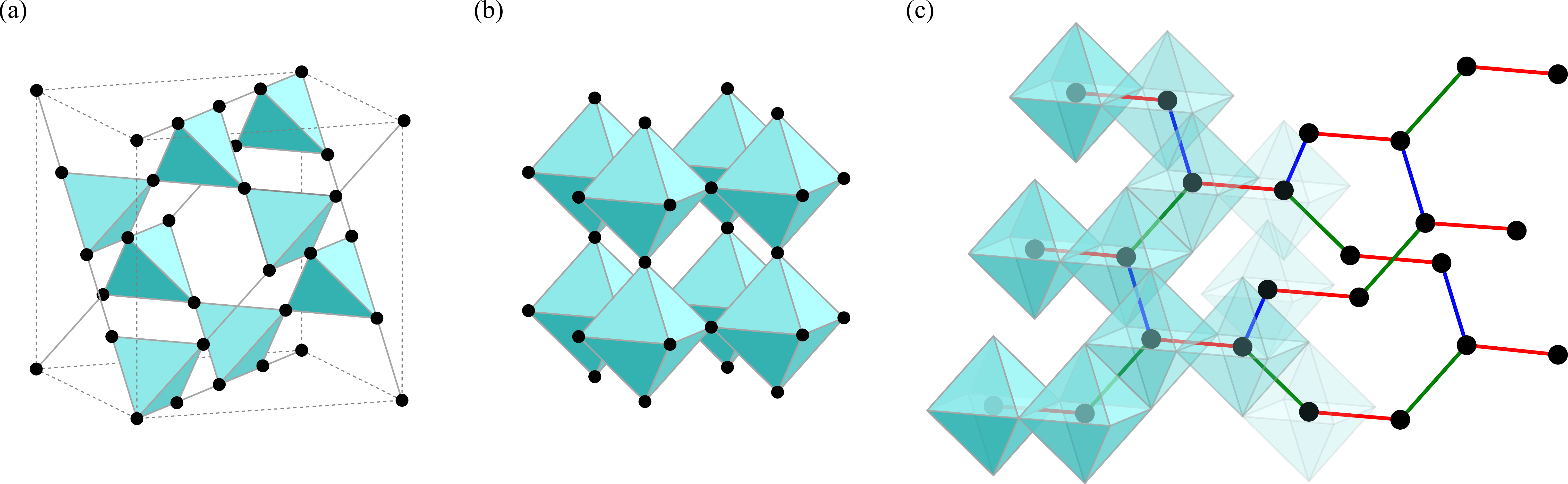}
    \caption{
Three-dimensional lattice architectures for constraint- and gauge-based
routes to quantum spin liquids.
(a) The pyrochlore lattice is formed from corner-sharing tetrahedra and
provides the standard three-dimensional platform for classical spin ice,
Coulomb phases, and quantum spin ice. The local tetrahedral constraint is
the microscopic origin of the two-in/two-out rule, emergent magnetic
charges, and the compact $U(1)$ gauge description.
(b) The octochlore lattice is formed from corner-sharing octahedra~\cite{Sklan-2013}. By
replacing tetrahedral units with octahedral ones, it enlarges the local
constraint structure and allows additional frustrated modes and coupling
channels. Recent work has proposed Ising-like octochlore magnets as a
platform for spin-ice-type Coulomb phases, fracton classical spin liquids,
and related constrained phases~\cite{Stern2026}.
(c) The hyperhoneycomb lattice, shown with the three Kitaev bond types
indicated by color, provides a canonical tricoordinated three-dimensional
setting for Kitaev physics. In the representation shown here, the
hyperhoneycomb graph threads the octochlore scaffold through a network of
bond- or edge-centered positions, suggesting a useful geometrical bridge
between corner-sharing cluster architectures and Kitaev-type
bond-anisotropic lattices. This panel emphasizes a complementary route to
three-dimensional spin liquids: not simplex constraints, but
bond-dependent exchange generating an emergent $\mathbb{Z}_2$ gauge
structure~\cite{Mandal-2009,Takayama2015PRL,OBrien-2016}.
}
    \label{fig:pyrochlore_octochlore}
\end{figure}

Seen in hindsight, two logically distinct obstacles were
often conflated in the search for quantum spin liquids.
The first is the destabilization of conventional ordered
states; the second is the actual availability of a liquid
phase once such ordering tendencies have been suppressed.
Early emphasis fell naturally on the first problem,
because reduced coordination and strong frustration
enhance fluctuations, and because underconstrained
lattices can support macroscopically degenerate manifolds
with no obvious symmetry-breaking outcome. This logic
was already visible in seminal work on pyrochlore
antiferromagnets, where weak connectivity and local
constraints were shown to produce disorder without
conventional ordering in both classical and quantum
settings \cite{MoessnerChalker1998PRL,Moessner-1998b,Canals-1998,Canals-2000,Moessner-2006}.
But solving this first problem did not by itself solve the
second: a large constrained manifold need not contain a
stable quantum liquid.

This constraint-based perspective is already visible at the level of lattice
architecture. The pyrochlore lattice, formed from corner-sharing tetrahedra
[Fig.~\ref{fig:pyrochlore_octochlore}(a)], is the canonical three-dimensional
setting in which local constraints give rise to classical spin ice, Coulomb
correlations, and, in the presence of quantum dynamics, $U(1)$ quantum spin
ice. It is therefore useful to ask whether similar principles can operate in
other three-dimensional networks of frustrated units.

A recent example is the octochlore lattice, a network of corner-sharing
octahedra [Fig.~\ref{fig:pyrochlore_octochlore}(b)]. Here, the elementary
frustrated unit is enlarged from a tetrahedron to an octahedron, allowing a
richer set of local modes and coupling patterns. Ising-like models on this
lattice have been proposed to realize several distinct constrained regimes,
including spin-ice-type Coulomb phases and a fracton classical spin liquid
with subdimensional excitations~\cite{Stern2026}. The comparison between
Figs.~\ref{fig:pyrochlore_octochlore}(a) and
\ref{fig:pyrochlore_octochlore}(b) illustrates a broader point that will recur
throughout this review: the pyrochlore is not simply an isolated geometrical
accident, but a paradigmatic example of how three-dimensional corner-sharing
cluster networks can stabilize constraint-dominated phases in frustrated
magnets.

The same architectural viewpoint also accommodates a complementary
three-dimensional route to spin-liquid physics. The hyperhoneycomb lattice
[Fig.~\ref{fig:pyrochlore_octochlore}(c)] is a tricoordinated
three-dimensional analogue of the honeycomb lattice and supports the canonical
Kitaev assignment of three inequivalent bond types. In the representation
shown in Fig.~\ref{fig:pyrochlore_octochlore}(c), the hyperhoneycomb graph can
be viewed as threading the octochlore scaffold through a network of bond- or
edge-centered positions, suggesting a useful geometrical bridge between
corner-sharing cluster architectures and Kitaev-type bond-anisotropic
lattices. Unlike the pyrochlore and octochlore examples, where the local
constraint originates from frustrated simplex or cluster units, the Kitaev
route generates an emergent $\mathbb{Z}_2$ gauge structure from
bond-dependent exchange on a tricoordinated network. The hyperhoneycomb
Kitaev model provides a canonical three-dimensional realization of this
principle~\cite{Mandal-2009,Takayama2015PRL,OBrien-2016}.

\begin{figure}[t]
    \centering
    \includegraphics[width=\textwidth]{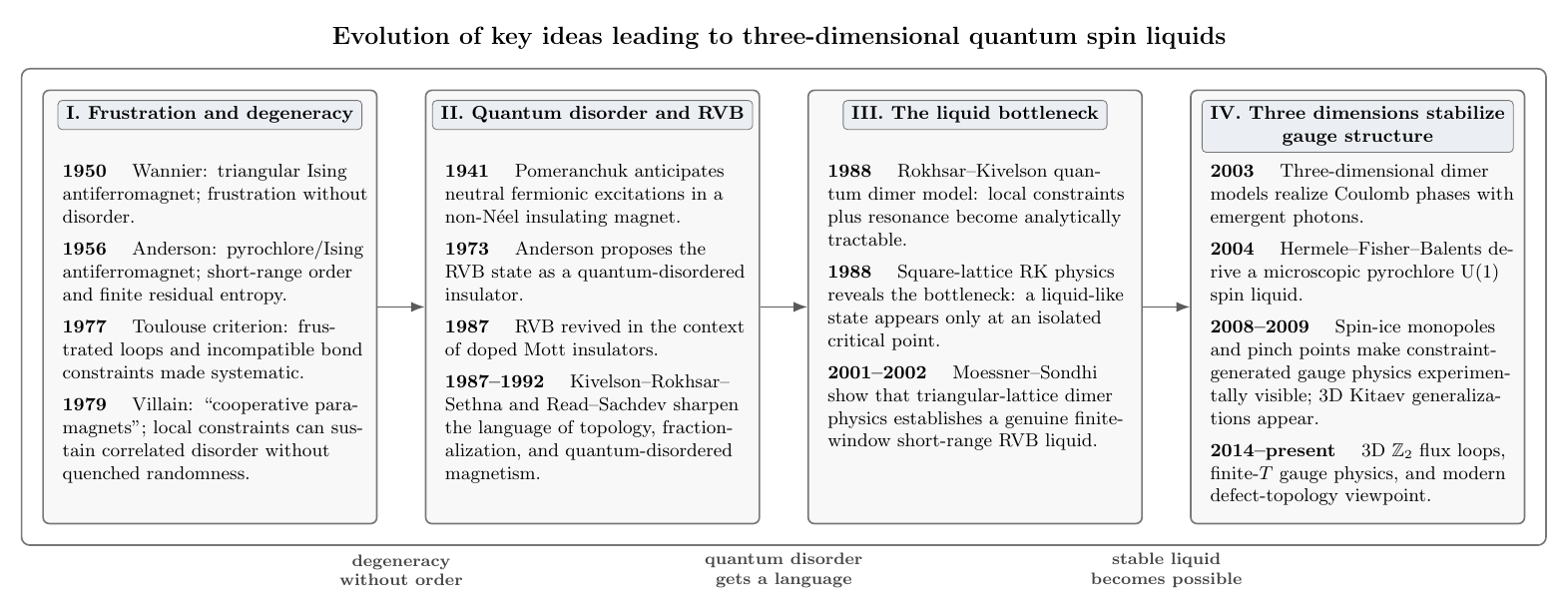}
    \caption{Historical sequence of ideas leading to modern three-dimensional quantum spin liquids. The progression runs from early recognition of frustration and macroscopic degeneracy, through Anderson’s RVB proposal and the emergence of topological and fractionalization language, to the Rokhsar--Kivelson dimer bottleneck and its resolution in finite-window liquid phases, culminating in the realization that three dimensions can stabilize emergent gauge structure. The figure emphasizes that 3D quantum spin liquids did not arise as an isolated conceptual jump, but as the continuation of a broader program linking constrained manifolds, resonance dynamics, and deconfinement.}
    \label{fig:timeline_history}
\end{figure}

The decisive conceptual advance on this second problem
came from Rokhsar--Kivelson-type quantum dimer models.
The original RK construction on the square lattice
provided an exactly tractable framework in which local
constraints, resonance dynamics, and equal-amplitude
superpositions could be handled analytically
\cite{Rokhsar-1988}. At the same time, it also revealed
the bottleneck: on the square lattice the liquid-like
state occurs only at an isolated critical point, while the
generic nearby phases are crystalline. As Moessner and
Sondhi put it, the short-range RVB liquid was still
``a phase in search of a Hamiltonian''
\cite{Moessner-2002}.

That barrier was overcome on the triangular lattice, but
for a reason that is more structural than a mere increase
of frustration. The crucial distinction is that the triangular
lattice is non-bipartite. On bipartite lattices such as the
square lattice, the dimer constraint naturally leads to an
emergent compact U(1)-type description in two spatial
dimensions, where instanton effects obstruct a stable
deconfined phase away from the RK point. On the
non-bipartite triangular lattice, by contrast, the effective
long-wavelength gauge structure is $\mathbb{Z}_2$, for which
a gapped deconfined phase is possible already in two
dimensions. Moessner and Sondhi showed that the
triangular-lattice quantum dimer model therefore hosts a
genuine short-range RVB phase over a finite interval of
parameters, with gapped collective modes and deconfined
spinons \cite{Moessner-2001}. The companion analysis of
the classical triangular-lattice dimer model further
clarified this distinction: the short-ranged dimer
correlations and deconfined monomers are consistent with
a stable $\mathbb{Z}_2$ liquid, rather than with a fine-tuned
U(1) critical point \cite{Fendley-2002,Henley-2004,Moessner-2002}.

Once this second obstacle had been removed, the move to
three dimensions was conceptually much milder than the
earlier step from ``frustration suppresses order'' to ``a
liquid phase actually exists.'' On the classical side,
cubic-lattice dimer models were shown to realize a
Coulomb phase with algebraic dipolar correlations
\cite{Huse-2003}. On the quantum side, three-dimensional quantum dimer models showed that bipartite lattices can support stable \(U(1)\) Coulomb liquids with emergent
photons and deconfined monomers, while non-bipartite three-dimensional dimer
models can realize gapped \(\mathbb{Z}_2\) RVB liquids~\cite{Moessner-2003,Huse-2003}. In parallel,
Hermele, Fisher, and Balents derived an explicit U(1)
spin liquid in the easy-axis pyrochlore antiferromagnet,
thereby embedding the same logic directly into a
microscopic frustrated spin Hamiltonian
\cite{HermeleFisherBalents2004PRB}. In this historical
flow, pyrochlore photons and related 3D U(1) liquids
are best viewed not as an isolated leap, but as a natural
continuation of the RK/dimer program once the availability of liquid phases had been established.

This historical development is summarized schematically in Fig.~\ref{fig:timeline_history}, which highlights the sequence of conceptual steps by which frustration, RVB ideas, dimer-liquid physics, and finally three-dimensional gauge structure converged into the modern understanding of 3D quantum spin liquids.

Figure~\ref{fig:timeline_history} places this development in historical context and also helps explain the origin of what might be called a tacit ``dimensional prejudice'': in the absence of an explicit liquid phase, three-dimensional frustrated magnets were often viewed through semiclassical frameworks in which harmonic fluctuations select order via order-by-disorder. Spin ice, pyrochlore antiferromagnets, dimer models, and lattice gauge mappings gradually shifted attention from fluctuation strength to the constraint algebra of the low-energy manifold and the infrared fate of the emergent gauge sector. The point is therefore not to overturn the usual dimensional intuition, but to sharpen it: dimensionality does not simply suppress quantum disorder; it reshapes the stability conditions of the gauge structure generated by local constraints.

A clean route begins with a classical spin
system whose ground-state sector is not a perturbation
of a single ordered configuration, but a macroscopically
degenerate constraint manifold. Spin ice provides the
canonical example: the local two-in/two-out rule
generates an extensive manifold with algebraic
correlations described by a divergence-free field
\cite{CastelnovoMoessnerSondhi2008Nature,GingrasMcClarty2014RoPP}.
The fundamental objects are not the spins themselves,
but the constraints and their defects. When quantum
dynamics—transverse exchange, multipolar matrix
elements, or ring-exchange processes—act within this
manifold, the long-wavelength theory is naturally an
emergent gauge theory
\cite{HermeleFisherBalents2004PRB,SavaryBalents2012PRL}.

Three-dimensional quantum spin liquids arise when:
(i) local constraints generate a macroscopically large low-energy sector,
(ii) dominant quantum processes act within this sector rather than selecting a unique ordered state, and
(iii) the resulting gauge structure avoids confinement.
Within this constraint-dominated setting, dimensionality can enhance
rather than suppress stability, because deconfinement of compact $U(1)$
gauge theory is more robust in three spatial dimensions, where it admits
a stable Coulomb phase over a finite region of coupling space. The microscopic problem is to find constraint manifolds and internal dynamics that frustrate symmetry-breaking selection at the outset.

The distinction between two and three dimensions can be understood most sharply at the level of gauge defects --- but \emph{this statement is gauge-structure dependent}. 
For compact U(1) gauge theories \emph{without} gapless matter, the key obstruction in $2+1$ dimensions is the proliferation of point-like instanton (monopole) events, which generically drive confinement in the infrared. 
In \(3+1\) dimensions, by contrast, the same compact U(1)
theory admits a stable deconfined Coulomb phase at weak
coupling, with a gapless photon and gapped magnetic
monopoles. The stability of this phase is controlled by the
finite monopole gap and the absence of monopole-instanton
proliferation that confines the \(2+1\)-dimensional theory~\cite{HermeleFisherBalents2004PRB,Fradkin-2013}. 
Thus, for compact U(1) gauge structure, dimensionality changes the infrared fate of the gauge sector.

For \(\mathbb{Z}_2\) gauge structure, deconfinement is possible in both two and three spatial dimensions at \(T=0\), but three dimensions introduce qualitatively new finite-temperature physics: the gauge-flux excitations are loop-like rather than point-like, enabling genuine finite-\(T\) transitions associated with flux-loop proliferation in several 3D constructions, including Kitaev-type models~\cite{Mandal-2009,Nasu-2014}. Accordingly, throughout this review we will be explicit about which stabilization statements are specific to compact U(1) gauge theories and which extend more broadly to other 3D QSL structures.

Connectivity alone is not decisive. Highly connected three-dimensional magnets without an underlying constraint manifold—such as conventional Heisenberg antiferromagnets on bipartite lattices—order at temperatures set by the dominant exchange scale despite large coordination numbers.
By contrast, in models where a local constraint enforces an emergent Gauss-law structure—such as quantum spin ice or quantum dimer models on bipartite three-dimensional lattices—a stable deconfined Coulomb phase can arise over a finite parameter regime.
The relevant variable is therefore not coordination number, but the existence and dynamical connectivity of a constraint sector.

Macroscopic degeneracy alone is not enough to produce a liquid upon quantization. The constrained sector must be locally connected by symmetry-allowed quantum processes, and these processes must appear at leading nontrivial order in perturbation theory. If constraint-preserving tunneling amplitudes arise only at parametrically higher order than symmetry-breaking perturbations, confinement or ordering can still intervene before gauge coherence develops. The ordering of energy scales is therefore as important as the existence of the constraint itself.

A further obstruction concerns the spatial structure of the leading constraint-preserving processes. Loop-like tunneling processes, as in quantum spin ice, naturally generate gauge-field kinetics because they connect the constrained manifold through extended closed moves. By contrast, leading processes that are too local or point-like may fail to produce an extended gauge dynamics: they can favor local resonances, leave the constrained manifold dynamically fragmented, or generate disconnected sectors rather than a coherent liquid. Thus the existence of a constraint manifold and symmetry-allowed perturbative processes is not by itself sufficient. What matters is whether the dominant processes establish an ergodic, loop-connected dynamics within the constrained sector.

This emphasis differs from platform-based reviews. Comprehensive accounts such as Refs.~\cite{SavaryBalents2016RoPP,GingrasMcClarty2014RoPP}
have emphasized material realizations, phenomenology, and gauge-theory
descriptions of established examples. Here we do not catalogue platforms or gauge structures. Instead, we trace the microscopic sequence required for deconfinement: local constraint enforcement, weak harmonic selection, constraint-preserving loop kinetics, and dimensional stabilization of the gauge sector. The emphasis
therefore shifts from cataloging examples to identifying the algebraic
and topological conditions under which deconfinement can persist in three
spatial dimensions.

\section{The False Dichotomy: Strong Fluctuations in 2D versus Classical Behaviour in 3D}

\subsection{The Traditional View}

The expectation that quantum spin liquids are predominantly low-dimensional has a clear origin.
Reduced coordination enhances zero-point motion, and continuous symmetries cannot be broken at finite temperature in two dimensions.
Many canonical spin-liquid candidates---triangular, kagome, and honeycomb magnets---are quasi-two-dimensional, reinforcing the intuition that strong fluctuations are the primary stabilizing ingredient~\cite{Balents2010Nature}.
Conversely, numerous three-dimensional magnets order readily at energy scales comparable to their exchange couplings, making it natural to view three-dimensional spin liquids as exceptional or finely tuned~\cite{SavaryBalents2016RoPP}.

At the most elementary level, frustration may be understood as the impossibility of simultaneously minimizing all pairwise interaction energies.
This idea already appears in classic examples such as Wannier's triangular-lattice Ising antiferromagnet~\cite{Wannier-1950}, and in classical spin systems it can be made more systematic through the Toulouse criterion~\cite{Toulouse-1977,Vannimenus-1977}, which identifies frustrated loops through an incompatible set of bond constraints.
Such local incompatibility naturally generates competing low-energy configurations and, in many important cases, an extensively degenerate ground-state manifold, which may support strong short-range correlations without developing long-range order, as emphasized in early work on frustrated lattices~\cite{Anderson-1956}.

A striking early indication that this intuition can fail came from experiments on highly frustrated magnets such as SrCr$_8$Ga$_4$O$_{19}$~\cite{Obradors-1988}, where the Curie--Weiss scale is very large, $|\theta|\sim 492\,\mathrm{K}$, yet no magnetic transition is observed down to about $4\,\mathrm{K}$.
This implies an anomalously large ratio $f=|\theta|/T_g\gtrsim 100$, far exceeding values typical of weakly frustrated magnets.
Such observations were later systematized by Ramirez through the introduction of the frustration parameter $f=|\theta_{\mathrm{CW}}|/T_{\mathrm{ord}}$ as a phenomenological measure of suppressed ordering scales in frustrated systems~\cite{Ramirez-1994}.
A large $f$, however, is only a symptom of frustrated ordering tendencies and not yet a diagnosis of a quantum spin liquid: systems with large frustration parameters may still freeze, develop short-range order, or select more conventional symmetry-broken states at lower temperatures. The essential question is therefore not simply why ordering is suppressed, but what replaces it.

A key step toward understanding the suppression of
ordering in frustrated magnets is the recognition that it
can arise from an extensive degeneracy of the ground-state
manifold, rooted in local constraints rather than symmetry.
This perspective has its roots in Villain's work on
cooperative paramagnets, where competing interactions
subject to local constraints generate macroscopically
degenerate states even in the absence of disorder
\cite{Villain-1977}. A simple way to see the origin of such
degeneracy is through Maxwell-type counting. For classical
Heisenberg spins of fixed length, each spin contributes two
continuous degrees of freedom, while each simplex imposes
vector constraints through the vanishing of the total spin
on that simplex. On corner-sharing simplex lattices these
constraints are not all independent, and both kagome and
pyrochlore antiferromagnets therefore possess extensive
classical ground-state manifolds. The distinction relevant
for the present review is not the mere existence of
degeneracy, but the dimensional setting and the associated
long-wavelength structure: the pyrochlore realizes a
three-dimensional constraint network whose Coulomb-phase
description and quantum dynamics lead naturally to compact
U(1) gauge theory in \(3+1\) dimensions.

This extensive degeneracy was already emphasized in early work by
Villain, who pointed out that frustrated antiferromagnets with a
large ground-state manifold behave as ``cooperative paramagnets,''
lacking long-range order even at zero temperature despite strong
interactions~\cite{Villain-1979}.
In such systems, the degeneracy is not accidental but reflects the
existence of a highly connected manifold of configurations satisfying
local constraints.
For nearest-neighbour Heisenberg models on kagome and pyrochlore
lattices, the Hamiltonian can be written as
$H \propto \sum_\alpha |\mathbf{L}_\alpha|^2$, where
$\mathbf{L}_\alpha = \sum_{i\in\alpha} \mathbf{S}_i$ is the total
spin on each simplex, so that ground states are defined by the local
constraint $\mathbf{L}_\alpha = 0$ on every simplex.

An important consequence, also stressed by Villain, is that such
highly degenerate systems are exceptionally sensitive to perturbations:
weak disorder, further-neighbour interactions, or thermal fluctuations
can lift the degeneracy and recast the system into qualitatively
new states, including spin-glass phases or, more generally, forms of
order emerging out of disorder~\cite{Villain-1980}. Closely related is the proliferation of metastable states within the degenerate manifold, reflecting a complex
energy landscape that further enhances sensitivity to perturbations and dynamical slowing. Thus, in these systems, the suppression of conventional order reflects an underconstrained structure, not simply large fluctuations.

This structural perspective also clarifies a key conceptual distinction
that emerged more gradually in the development of the field.
Early studies of frustrated magnets established that local constraints
can suppress conventional order by generating macroscopically
degenerate manifolds, leading to cooperative paramagnetic behaviour
with strong correlations but no symmetry breaking.
However, subsequent work---particularly on quantum dimer models---showed
that the absence of order does not, by itself, imply the presence of a
quantum spin liquid.
In fact, certain frustrated models can be mapped directly onto dimer
coverings, making explicit the connection between local constraints,
extensive degeneracy, and emergent gauge structure~\cite{Villain-1977}.
While many highly frustrated systems remain disordered down to low
temperatures, only in special cases do quantum fluctuations reshape
the constrained manifold into a stable deconfined phase. Frustration therefore answers only the first question—why order is suppressed. It leaves open the harder question: when is a liquid phase actually stabilized? Three dimensions play a crucial role precisely at this second stage, since the topology and thermodynamics of gauge defects can permit deconfinement to persist over finite parameter regimes, in contrast to
the generic confinement of compact gauge theories in two dimensions.

The relevant distinction is whether the suppressed ordering reflects the presence of a constraint-generated manifold with emergent gauge structure, or merely competition among nearly degenerate classical states.
The former case provides the most direct route to a stable gauge-theoretic quantum spin liquid. Other routes are possible, but then the mechanism of stabilization must be identified separately, rather than inferred from frustration or suppressed ordering alone.

This reasoning implicitly assumes that magnetic order is selected from a small set of nearly classical competing states.
In systems with a unique or weakly degenerate classical ground state, fluctuations either renormalize or destabilize that order.
Under those circumstances, increasing dimensionality indeed favors classical behavior. More generally, frustrated systems can exhibit partial disorder or reentrant behavior, in which ordered and disordered degrees of freedom coexist, underscoring that the suppression of order is not a binary phenomenon but can involve highly structured intermediate regimes~\cite{Liebmann1986}. Even within conventional ordering scenarios, however, the stabilizing effect of dimensional crossover is weaker than often assumed.

In quasi-two-dimensional antiferromagnets, long-range order induced by weak interlayer coupling sets in only once the in-plane correlation length becomes exponentially large. As a result, in the renormalized classical regime the ordering temperature is parametrically suppressed and behaves as~\cite{Chakravarty-1989,Yasuda-2005}
\begin{equation}
T_c \sim \frac{4\pi \rho_s}{\ln\!\left(J_\parallel/J_\perp\right) + b}\,,
\end{equation}
where $\rho_s$ is the in-plane spin stiffness and $b$ is a nonuniversal constant. Subleading corrections in $1/\ln(J_\parallel/J_\perp)$ and model-dependent numerical factors are omitted. This expression should therefore be understood as a parametric statement about fragility rather than as a universal formula. Dimensionality thus regulates divergent correlations; it does not, by itself, automatically generate a large ordering scale in the absence of a unique classical ground state.

These considerations weaken the simplistic association between higher dimension and classical behaviour. But they do not yet explain why fully three-dimensional frustrated magnets can avoid ordering altogether. For that, one must look not to the magnitude of fluctuations alone, but to the structure of the low-energy manifold itself.

When dominant interactions impose local constraints—whether geometric, algebraic, or symmetry-enforced—the system does not fluctuate around a unique classical configuration. Instead, it resides within an extensively degenerate manifold whose structure is more naturally described in gauge-theoretic variables than in terms of conventional order parameters. While local constraint manifolds typically originate from frustrated exchange interactions, not all frustrated systems generate algebraic Gauss-law–type structures. The crucial distinction is whether the degeneracy can be expressed as a local conservation law acting on simplex units, rather than as a competition among distinct global ordering tendencies. Only in the former case does the low-energy sector lend itself naturally to a gauge-theoretic description.

The central question is therefore not how strong quantum fluctuations must be in three dimensions, but whether the mechanisms that normally select order are rendered ineffective before such selection can occur.

\subsection{Reframing the Question}

This tension between fluctuation-selected order and quantum-disordered
states was already visible in early large-$N$ treatments of kagome and
triangular antiferromagnets~\cite{Sachdev-1992}. In constraint-dominated
systems, the relevant degrees of freedom are collective: the manifold is
organized by local conservation laws and emergent gauge structure rather
than by proximity to a single ordered pattern~\cite{GingrasMcClarty2014RoPP}.
Quantum dynamics that respect these constraints generate tunnelling within
the manifold, and in three dimensions the associated gauge theory can remain
deconfined~\cite{HermeleFisherBalents2004PRB,SavaryBalents2012PRL}.
Neutron scattering reveals spin correlations consistent with such emergent
gauge structure, most prominently through pinch-point singularities associated
with a divergence-free constraint~\cite{Benton2012PRB}.

Dimensionality enters through the infrared stability of the
emergent gauge sector. Three dimensions permit stable
Coulomb phases of compact U(1) gauge fields and loop-like
flux excitations in \(\mathbb{Z}_2\) spin liquids
\cite{HermeleFisherBalents2004PRB,KimchiVishwanath2014PRB}. This is the setting of the constraint-first mechanism discussed below.

The criterion also has limits. We do not claim that every macroscopically constrained three-dimensional manifold quantizes into a spin liquid, nor that an increased density of loops or higher connectivity automatically promotes liquid behavior. 
Stabilization requires that constraint-preserving dynamics arise at leading nontrivial order in perturbation theory and that gauge-breaking perturbations remain parametrically weak. Moreover, the framework does not resolve materials-specific ambiguities where disorder, multipolar mixing, or additional couplings dominate the low-energy physics. 
The constraint-first perspective is therefore a structural criterion for identifying when three-dimensional quantum spin liquids can arise, not a guarantee that they must.

The constraint principle extends beyond spin-ice--type Coulomb phases.
Projector constructions such as the AKLT family provide an instructive limit in which local Hamiltonians enforce valence-bond constraints exactly, stabilizing fully gapped paramagnets even on highly connected three-dimensional lattices \cite{Affleck1987PRL,Arovas1988PRL}. In these models, local projector terms penalize configurations outside a constrained valence-bond manifold, thereby stabilizing a nonmagnetic ground state by construction. Importantly, three-dimensional realizations show that such
constraint-enforced states can remain quantum disordered:
depending on lattice geometry and spin magnitude, frustrated
3D AKLT constructions can support nonmagnetic ground
states~\cite{Parameswaran2009PRB}. These states are short-range entangled and realize symmetry-protected topological (SPT) phases rather than intrinsic topological order. Nevertheless, they demonstrate that in three dimensions robust quantum-disordered phases can arise purely from local constraint enforcement.
From this perspective, fractionalized quantum spin liquids may be viewed as the long-range--entangled extension of the same underlying logic: when the constrained manifold admits an emergent gauge-field description
and supports deconfined excitations, intrinsic topological order replaces SPT structure.
This comparison separates two forms of constraint-enforced quantum disorder: short-range-entangled projector paramagnets and long-range-entangled gauge phases.

More broadly, theoretical constructions such as the Hubbard model on the pyrochlore lattice demonstrate that three-dimensional constraint manifolds can host quantum spin liquids without relying on extreme fluctuation regimes \cite{NormandNussinov2014PRL}.
Experimentally, frustrated three-dimensional networks such as PbCuTe$_2$O$_6$ and K$_2$Ni$_2$(SO$_4$)$_3$ illustrate how connectivity and competing exchanges can generate extended liquid-like regimes rather than conventional order \cite{Chillal2020NatComm,Gonzalez2024NatComm,Zivkovic-2021}.

Exactly solvable $\mathbb{Z}_{2}$ models provide a second route. The paradigmatic exactly solvable realization is Kitaev's honeycomb model, in which spins fractionalize into itinerant Majorana fermions coupled to a static emergent
\(\mathbb{Z}_2\) gauge field~\cite{Kitaev-2006}. Exactly solvable three-dimensional Kitaev constructions make this dimensional distinction explicit, exhibiting deconfined gauge structure and genuine finite-temperature phase transitions---associated with gauge-flux ordering---that are forbidden in two-dimensional $\mathbb{Z}_2$ gauge systems~\cite{Nasu-2014,Takayama2015PRL,Trebst2022PhyRep}.
In three dimensions, explicit solvable generalizations beyond the original honeycomb model exist. 
Mandal and Surendran introduced a fully solvable 3D Kitaev-type model demonstrating stable $\mathbb{Z}_2$ deconfinement and both gapped and gapless spin-liquid phases \cite{Mandal-2009}. 
Thus, three-dimensional $\mathbb{Z}_{2}$ gauge liquids are controlled microscopic phases, not just heuristic extensions of two-dimensional models.

In three dimensions, lattice geometry also shapes the structure of the gapless Majorana sector. For elementary tricoordinated 3D lattices---including members of the harmonic-honeycomb family---the itinerant Majorana spectrum can realize nodal lines, extended Majorana Fermi surfaces, or Weyl nodes when time-reversal symmetry is broken. 
This classification provides a systematic taxonomy of gapless $\mathbb{Z}_2$ spin liquids in three dimensions and directly links lattice connectivity to low-energy quasiparticle topology \cite{OBrien-2016}. 
Unlike in two dimensions, where Dirac cones are generic, three dimensions allow stable codimension-one and codimension-two nodal manifolds protected by lattice symmetries.

A further subtlety arises when the gauge sector itself becomes frustrated.
In certain generalized three-dimensional Kitaev models, the static $\mathbb{Z}_2$
flux configuration is not uniquely selected but instead exhibits competing
gauge patterns, generating additional energy scales and finite-temperature
transitions within the gauge sector itself~\cite{Eschmann-2019}.
Here, the competition lies not only in the Majorana sector but also in the emergent gauge structure.

The gauge sector also changes in three dimensions. Flux excitations are loop-like rather than point-like, and their proliferation can support finite-temperature phase transitions. 
Large-scale sign-free simulations on the hyperhoneycomb Kitaev lattice demonstrated such a finite-$T$ transition associated with flux-loop condensation \cite{Nasu-2014}. 
Closely related physics was identified on the hyperoctagon lattice, where a thermodynamic ``liquid--gas'' transition of flux loops further reinforced that finite-temperature gauge phenomena are a recurrent feature of three-dimensional Kitaev systems~\cite{Mishchenko-2017}. This loop-gauge physics is one of the sharpest qualitative distinctions between 2D and 3D $\mathbb{Z}_2$ spin liquids.

A useful feature of exactly solvable 3D Kitaev models is the controlled calculation of dynamical response functions. 
The dynamical spin structure factor (``Majorana spectroscopy'') has been computed for hyperhoneycomb and related lattices, providing concrete predictions for neutron scattering signatures of fractionalization \cite{Smith-2015,Smith-2016}. 
Complementary Raman response calculations tailored to the 3D harmonic-honeycomb family further establish polarization-resolved fingerprints of fractionalized excitations \cite{Perreault-2015}. 
Because these responses can be obtained without uncontrolled approximations, they provide controlled benchmarks linking gauge-theory phenomenology directly to experiment.

On the materials side, \(\beta\)- and \(\gamma\)-\(\mathrm{Li_2IrO_3}\) established canonical three-dimensional Kitaev-relevant platforms. Although these materials ultimately develop magnetic order, their bond-dependent interactions and broad excitation continua above ordering temperatures motivate their interpretation as proximate Kitaev magnets. Polarization-resolved Raman measurements on $\beta$- and $\gamma$-$\mathrm{Li_2IrO_3}$ report broad continua consistent with fractionalized fermionic excitations in a Kitaev-dominated regime \cite{Glamazda-2016}. 
Thus, even when long-range order intervenes, three-dimensional Kitaev magnets demonstrate that symmetry-allowed bond anisotropies can stabilize extended fractionalized regimes in realistic compounds.

These examples show that three-dimensional quantum
spin liquids can arise through two complementary routes:
(i) constraint-first Coulomb phases with emergent \(U(1)\) gauge structure
and gapless photons, and
(ii) exactly solvable or proximate \(\mathbb{Z}_2\) gauge liquids with
itinerant Majorana fermions and loop-like gauge-flux excitations.
The two routes exploit different three-dimensional features: the stability
of compact \(U(1)\) Coulomb phases in \(3+1\) dimensions on the one hand,
and the topology of \(\mathbb{Z}_2\) flux loops, Majorana nodal structures,
and finite-temperature gauge-sector transitions on the other.


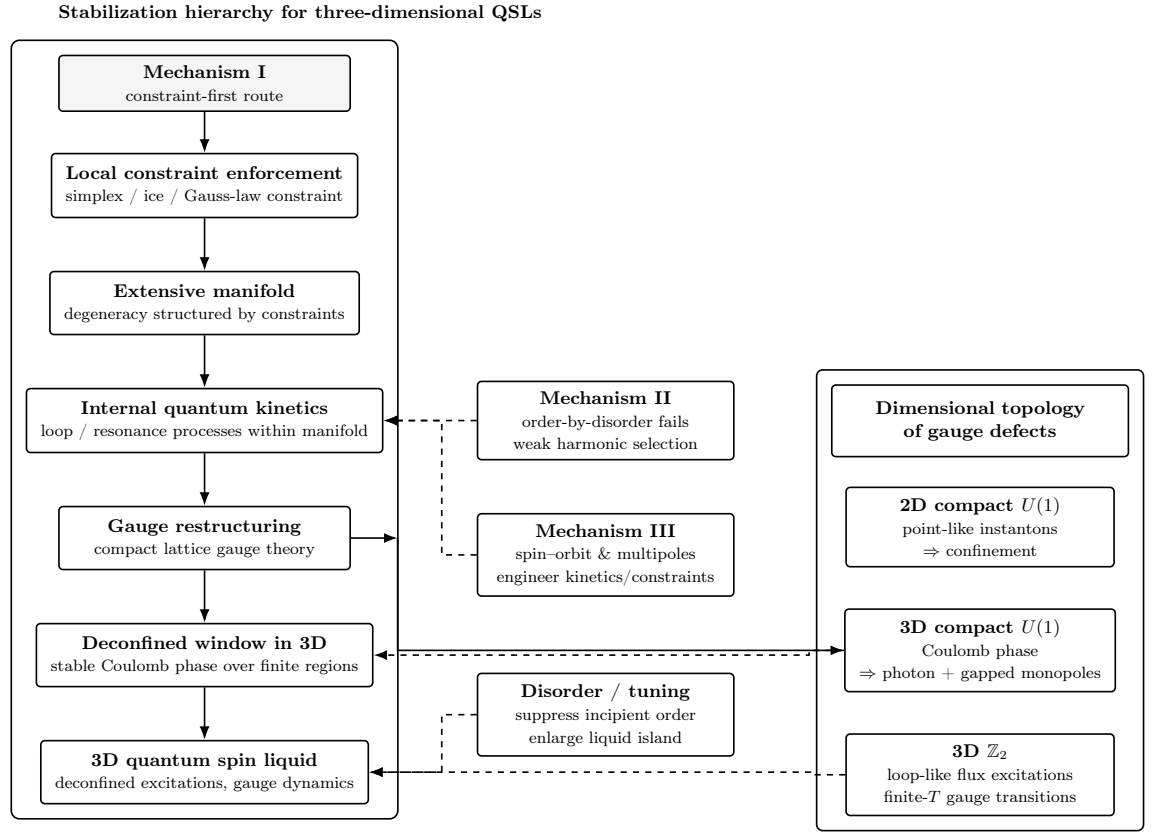
\begin{figure}[t]
\centering
\resizebox{0.98\linewidth}{!}{%
\begin{tikzpicture}[
  font=\small,
  >=Latex,
  box/.style={
    draw, rounded corners=2pt, thick,
    align=center, inner xsep=7pt, inner ysep=6pt,
    fill=white
  },
  sbox/.style={
    draw, rounded corners=2pt, thick,
    align=center, inner xsep=7pt, inner ysep=5pt,
    fill=white
  },
  panel/.style={
    draw, rounded corners=3pt, thick,
    inner xsep=8pt, inner ysep=8pt,
    fill=white
  },
  arr/.style={->, thick},
  harr/.style={->, thick, dashed},
  title/.style={font=\bfseries\small}
]

\node[sbox, minimum width=5.0cm, fill=gray!8] (m1)
{\textbf{Mechanism I}\\
\footnotesize constraint-first route};

\node[box, below=7mm of m1, minimum width=5.0cm] (c1)
{\textbf{Local constraint enforcement}\\
\footnotesize simplex / ice / Gauss-law constraint};

\node[box, below=9mm of c1, minimum width=5.0cm] (c2)
{\textbf{Extensive manifold}\\
\footnotesize degeneracy structured by constraints};

\node[box, below=9mm of c2, minimum width=5.0cm] (c3)
{\textbf{Internal quantum kinetics}\\
\footnotesize loop / resonance processes within manifold};

\node[box, below=9mm of c3, minimum width=5.0cm] (c4)
{\textbf{Gauge restructuring}\\
\footnotesize compact lattice gauge theory};

\node[box, below=9mm of c4, minimum width=5.0cm] (c5)
{\textbf{Deconfined window in 3D}\\
\footnotesize stable Coulomb phase over finite regions};

\node[box, below=9mm of c5, minimum width=5.0cm] (c6)
{\textbf{3D quantum spin liquid}\\
\footnotesize deconfined excitations, gauge dynamics};

\draw[arr] (m1) -- (c1);
\draw[arr] (c1) -- (c2);
\draw[arr] (c2) -- (c3);
\draw[arr] (c3) -- (c4);
\draw[arr] (c4) -- (c5);
\draw[arr] (c5) -- (c6);

\begin{scope}[on background layer]
  \node[
    draw,
    rounded corners=4pt,
    thick,
    inner sep=7pt,
    fit=(m1)(c1)(c2)(c3)(c4)(c5)(c6)
  ] (m1frame) {};
\end{scope}

\node[sbox, right=16mm of c3, minimum width=4.4cm] (m2)
{\textbf{Mechanism II}\\
\footnotesize order-by-disorder fails\\
\footnotesize weak harmonic selection};

\node[sbox, below=9mm of m2, minimum width=4.4cm] (m3)
{\textbf{Mechanism III}\\
\footnotesize spin--orbit \& multipoles\\
\footnotesize engineer kinetics/constraints};

\node[sbox, below=13mm of m3, minimum width=4.4cm] (md)
{\textbf{Disorder / tuning}\\
\footnotesize suppress incipient order\\
\footnotesize enlarge liquid island};

\draw[harr] (m2.west) -- ++(-6mm,0) |- (c3.east);
\draw[harr] (m3.west) -- ++(-6mm,0) |- (c3.east);
\draw[harr] (md.west) -- ++(-6mm,0) |- (c6.east);


\node[panel, right=18mm of m2, minimum width=4.8cm] (topPanel) {};

\node[sbox, align=center, text width=4.6cm, inner ysep=8pt]
  (topTitle) at ($(topPanel.north)+(0,-3mm)$)
  {\bf Dimensional topology\\of gauge defects};

\node[sbox, anchor=north, minimum width=4.6cm]
  (t2d) at ($(topTitle.south)+(0,-5mm)$)
  {\textbf{2D compact \(U(1)\)}\\
   \footnotesize point-like instantons\\
   \footnotesize \(\Rightarrow\) confinement};

\node[sbox, below=7mm of t2d, minimum width=4.6cm]
  (t3d)
  {\textbf{3D compact \(U(1)\)}\\
   \footnotesize Coulomb phase\\
   \footnotesize \(\Rightarrow\) photon + gapped monopoles};

\node[sbox, below=7mm of t3d, minimum width=4.6cm]
  (tz2)
  {\textbf{3D \(\mathbb{Z}_2\)}\\
   \footnotesize loop-like flux excitations\\
   \footnotesize finite-\(T\) gauge transitions};

\begin{scope}[on background layer]
  \node[panel, fit=(topTitle)(t2d)(t3d)(tz2), inner sep=7pt] (topFit) {};
\end{scope}

\coordinate (junc) at ($(c4.east)+(8mm,0)$);
\draw[arr] (c4.east) -- (junc);
\draw[arr] (junc) |- (t3d.west);

\draw[harr] (t3d.west) -- ++(-6mm,0) |- (c5.east);

\draw[harr] (tz2.west) -- ++(-6mm,0) |- (c6.east);

\node[title, inner sep=0pt, anchor=west] (ttl)
  at ($(m1.north west)+(0,7mm)$)
{Stabilization hierarchy for three-dimensional QSLs};

\path[use as bounding box]
  (ttl.west |- current bounding box.south)
  rectangle
  (current bounding box.north east);

\end{tikzpicture}%
}

\caption{
Mechanism I: constraint-first stabilization hierarchy for
three-dimensional QSLs. Dominant interactions enforce local constraints,
generating an extensive manifold; symmetry-allowed transverse or multipolar
processes generate dynamics within that manifold; and the low-energy theory
may be recast as a compact lattice gauge theory. The role of dimensionality
is gauge-structure dependent. For compact \(U(1)\) gauge theories without
gapless matter, monopole instantons generically confine the theory in
\(2+1\) dimensions, whereas in \(3+1\) dimensions a stable weak-coupling
Coulomb phase exists, characterized by a gapless photon and gapped magnetic
monopoles. For \(\mathbb{Z}_2\) gauge theories, deconfinement is possible
already at zero temperature in two spatial dimensions; the distinctive
three-dimensional feature is instead the loop-like character of gauge-flux
excitations and the possibility of finite-temperature gauge-sector
transitions. The additional mechanisms discussed below act as amplifiers by
suppressing premature order selection and by helping generate
constraint-preserving kinetics.
}
\label{fig:hierarchy}
\end{figure}

\section{Mechanism I: Constraint-First Route and Why Three Dimensions Help}

\subsection{Classical Constraint Manifolds and Coulomb Phases}

The constraint-first route to three-dimensional quantum spin liquids begins in the classical limit.
In contrast to magnets whose ground states are perturbations of a single symmetry-breaking configuration, certain frustrated lattices possess an extensively degenerate manifold defined by local constraints.
The pyrochlore Heisenberg antiferromagnet provides a canonical example. For nearest-neighbor exchange
\begin{equation}
H = J \sum_{\langle ij \rangle} \mathbf S_i \cdot \mathbf S_j ,
\end{equation}
the classical ground-state condition can be rewritten as a simplex constraint
\begin{equation}
\sum_{i \in \text{tet}} \mathbf S_i = 0 ,
\end{equation}
on every tetrahedron. Minimizing the nearest-neighbor exchange enforces the condition that the total spin on each tetrahedron vanishes.
This local zero-divergence constraint generates a macroscopically degenerate manifold whose correlations are algebraic and governed at long wavelengths by an emergent divergence-free field \cite{Isakov2004PRL,Henley2010ARCMP}. 

Spin ice sharpens this structure further. The two-in/two-out
rule on each tetrahedron imposes a discrete Gauss-law
constraint, producing a classical Coulomb phase
characterized by dipolar correlations and pinch-point
singularities in the structure factor. Castelnovo, Moessner,
and Sondhi recast violations of the ice rule as
fractionalized magnetic charges interacting via an emergent
Coulomb law, thereby establishing the monopole language
that now underpins the modern interpretation of spin
ice~\cite{CastelnovoMoessnerSondhi2008Nature}. This
language provided a transparent bridge between
gauge-theory concepts and experiment, enabling direct
interpretation of neutron-scattering and thermodynamic
data in terms of deconfined gauge charges rather than
conventional spin flips. Direct experimental confirmation
of the Coulomb phase, including the characteristic
pinch-point singularities, was obtained through polarized
neutron-scattering measurements on Ho$_2$Ti$_2$O$_7$,
demonstrating near-ideal divergence-free correlations in a
real material~\cite{Fennell-2009}. The broader
phenomenology of magnetic pyrochlore oxides and their
Coulomb phases has been comprehensively reviewed in
Ref.~\cite{GingrasMcClarty2014RoPP}.

Such constraint manifolds already encode an emergent gauge structure at the
classical level. In spin ice, the allowed configurations can be mapped onto
a coarse-grained divergence-free field, while violations of the ice rule act
as effective magnetic charges. This provides the natural starting point for
the quantum spin-ice construction, where transverse exchange generates
coherent dynamics within the constrained manifold.

More explicitly, in spin ice one assigns to each pyrochlore site \(i\) a
local Ising axis \(\hat z_i\), corresponding to one of the four inequivalent
local \(\langle 111\rangle\) directions of the pyrochlore sublattices. A
coarse-grained field on the dual diamond lattice may then be defined as
\begin{equation}
{\bf B}({\bf r}) \propto \sum_{i\in {\rm tet}} \hat z_i S_i^z ,
\end{equation}
where \(S_i^z\) is the Ising component along the local axis \(\hat z_i\).
The two-in/two-out ice rule is the lattice divergence-free condition whose
coarse-grained continuum form is
\begin{equation}
\nabla\cdot{\bf B}=0 .
\end{equation}

For the classical Heisenberg pyrochlore antiferromagnet, an analogous
Coulomb-phase description involves three independent divergence-free fields,
one for each spin component. Quantum spin ice, however, is most directly
realized in anisotropic Ising or XXZ-type pyrochlore models, where a
dominant local Ising exchange enforces the ice-rule manifold and transverse
exchange generates ring dynamics within it.

In the continuum limit, the divergence-free constraint identifies the
classical Coulomb phase with an emergent Maxwell theory, where fluctuations
of \({\bf B}\) correspond to transverse gauge modes. Constraint correlations
alone, however, are not sufficient to establish a quantum spin liquid. A
genuinely quantum regime requires coherent dynamics within the constrained
sector, manifested for example in a well-defined emergent photon mode and
quantum-renormalized pinch-point correlations persisting to low temperature
without freezing.

Figure~\ref{fig:hierarchy} collects the steps of the constraint-first route, which serves as a roadmap for the mechanisms developed in the following sections. The central column represents Mechanism I, the constraint-first route: a local constraint generates an extensive manifold, quantum kinetics act within that manifold, and the
resulting low-energy theory can enter a deconfined gauge regime in three
dimensions.

\subsection{Quantum Dynamics within Constrained Manifolds}

\begin{figure}
\includegraphics[width = 0.88\textwidth]{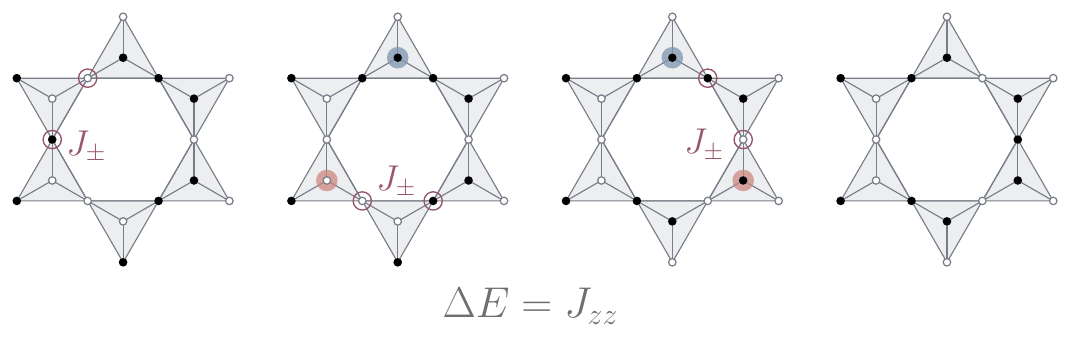}
\caption{
Microscopic origin of loop dynamics in quantum spin ice. In the XXZ limit,
the dominant \(J_{zz}\) term enforces the ice-rule manifold, while a
transverse exchange \(J_{\pm}\) creates virtual pairs of defects violating
the ice rule at an energy cost \(\Delta E\sim J_{zz}\). Through a sequence
of virtual processes, these defects propagate around a closed hexagonal loop
and recombine, generating an effective ring-exchange process of order
\(K\sim J_{\pm}^3/J_{zz}^2\) within the constrained manifold. This illustrates
how quantum dynamics act within the ice-rule sector to produce emergent
gauge kinetics.
}
\label{fig:qsi_tunneling}
\end{figure}

Quantum mechanics does not necessarily destabilize these manifolds.
Instead, it can endow them with dynamics. In spin ice and related models,
transverse exchange terms or higher-order processes generate tunneling
between distinct constraint-satisfying configurations. In perturbation theory
about the classical limit, such processes often produce effective
ring-exchange terms that act within the constrained subspace
\cite{HermeleFisherBalents2004PRB,Shannon2012PRL}.

For concreteness, consider the standard quantum spin-ice limit of an
anisotropic nearest-neighbor exchange model,
\begin{equation}
H_{\rm XXZ}
=
J_{zz}\sum_{\langle ij\rangle} S_i^{z_i} S_j^{z_j}
-
J_{\pm}\sum_{\langle ij\rangle}
\left(S_i^+S_j^-+S_i^-S_j^+\right)
+\cdots ,
\end{equation}
where \(z_i\) denotes the local \(\langle 111\rangle\) axis on sublattice \(i\),
and \(S_i^\pm\) are defined in the corresponding local transverse frame. The dominant \(J_{zz}\) term enforces
the ice-rule manifold, while \(J_{\pm}\) generates virtual excursions out of
this manifold. At order \(n\) in degenerate perturbation theory about the
Ising limit, the resulting constraint-preserving loop amplitude scales
parametrically as
\begin{equation}
K \sim \frac{J_{\pm}^{\,n}}{J_{zz}^{\,n-1}} .
\end{equation}
For the elementary hexagonal ring exchange of quantum spin ice, \(n=3\), so
\(K\) is of order \(J_{\pm}^{3}/J_{zz}^{2}\), up to a numerical prefactor.

A representative microscopic process underlying such loop dynamics is
illustrated in Fig.~\ref{fig:qsi_tunneling}. Starting from an ice-rule
configuration, a transverse exchange term \(J_{\pm}\) locally violates the
constraint by creating a pair of defects with an energy cost
\(\Delta E\sim J_{zz}\). Subsequent virtual processes propagate these defects
around a closed loop before annihilating them, returning the system to the
constrained manifold. At lowest nontrivial order, this generates an effective
ring-exchange process acting on hexagonal loops of the pyrochlore lattice,
providing the elementary kinetic term within the quantum spin-ice manifold.

The resulting quantum dynamics can be formulated as a compact lattice gauge
theory. For spin-\(1/2\) pyrochlore systems with dominant Ising exchange,
transverse terms generate a quantum spin-ice Hamiltonian whose low-energy
description is a compact \(U(1)\) gauge theory with gapless photon
excitations and gapped electric and magnetic charges
\cite{HermeleFisherBalents2004PRB,SavaryBalents2012PRL}. Multipolar
interactions in rare-earth pyrochlores provide another microscopic
realization of the same constraint-first mechanism: non-Ising matrix elements
can act within, or virtually connect, the constrained manifold and generate
coherent tunneling dynamics~\cite{GingrasMcClarty2014RoPP}.

At this stage the dimensional distinction becomes important. For compact
\(U(1)\) gauge theories without gapless matter, monopole-instanton
proliferation generically confines the would-be deconfined phase in
\(2+1\) dimensions. In \(3+1\) dimensions, by contrast, compact \(U(1)\)
gauge theory admits a stable weak-coupling Coulomb phase with a gapless
photon and gapped magnetic monopoles
\cite{Fradkin-2013,HermeleFisherBalents2004PRB}. For \(\mathbb{Z}_2\)
gauge theories, deconfinement is already possible at zero temperature in two
spatial dimensions; the distinctive three-dimensional feature is instead
that gauge-flux excitations are loop-like and can support finite-temperature
gauge-sector transitions
\cite{Wegner-1971,Kitaev-2006,Mandal-2009,Nasu-2014}.

The key structural point is that the dominant quantum processes preserve the
local constraints to leading order. They therefore generate dynamics within a
pre-existing manifold rather than selecting a symmetry-breaking state. The
gauge structure is not imposed by hand at the microscopic level; it emerges
in the coarse-grained description of the constrained manifold, where the
physical spin configurations can be represented by fields satisfying a
Gauss-law constraint.

\subsection{Deconfinement and Emergent Gauge Fields in Three Dimensions}

For compact \(U(1)\) gauge theory without gapless matter, \(3+1\) dimensions
admit a stable weak-coupling Coulomb regime. The effective description
obtained above therefore has the long-wavelength Maxwell form
\begin{equation}
\mathcal L = \frac{1}{2e^2} \left( \mathbf E^2 - c^2 \mathbf B^2 \right),
\end{equation}
where \(\mathbf E\) and \(\mathbf B\) are emergent electric and magnetic
fields defined in terms of the coarse-grained gauge potential, with emergent
photon velocity \(c\) and gauge coupling \(e\) set by microscopic parameters.
This distinction is not a quantitative enhancement of fluctuations but a
qualitative change in gauge-theory stability.

Once microscopic dynamics generate the appropriate compact \(U(1)\) gauge
structure, this weak-coupling regime corresponds to a deconfined Coulombic
quantum spin liquid. Monopole proliferation is suppressed, the photon remains
deconfined, and neutron-scattering signatures include quantum-broadened
pinch points together with linearly dispersing photon modes~\cite{Benton2012PRB}.
The constraint-first route therefore explains both how emergent gauge
theories arise in frustrated magnets and why their deconfined regime can
persist in three spatial dimensions.

The existence of an extensively degenerate classical manifold does not guarantee
that quantum dynamics will stabilize it as a liquid. Order-by-disorder can
intervene before coherent gauge dynamics develop, lifting the degeneracy and
selecting conventional order. We therefore turn to the conditions under which
this selection is weak. This is the subject of Sec.~\ref{sec:mechII}.

\section{Mechanism II: When Order-by-Disorder Fails}
\label{sec:mechII}
\subsection{Macroscopic Degeneracy and Ineffective Harmonic Selection}

Order-by-disorder (OBD) is often invoked as the generic resolution of frustration: a classically degenerate manifold is expected to be lifted once thermal or quantum fluctuations generate an effective potential on the manifold \cite{Henley1989PRL}.
This logic is reliable when (i) the ground-state manifold can be parametrized by a small number of collective coordinates, and (ii) the harmonic fluctuation spectrum varies appreciably across the manifold, so that the zero-point or entropic free energy produces an extensive selection. Neither premise is guaranteed in three-dimensional constraint-dominated magnets. 

We distinguish between continuous manifolds parametrized by soft collective modes, as in the classical Heisenberg pyrochlore antiferromagnet~\cite{Reimers-1992,Moessner-1998b}, and discrete constraint manifolds such as spin ice, where allowed configurations are defined by local ice-rule constraints~\cite{Harris-1997,Bramwell-2001}. In the former, low-energy fluctuations can be large because extended soft modes exist across the manifold; in the latter, configurations are locally stable and connected primarily by constraint-preserving rearrangements. The inefficiency of harmonic selection discussed here is most transparent in the latter setting, where quadratic fluctuation spectra are nearly identical across locally related configurations~\cite{Henley1989PRL,Henley2006PRL}.

The key distinction is that the degeneracy in constraint-dominated systems is locally connected, in a gauge-like manner, rather than parametrized by a small set of global collective coordinates. In manifolds of the latter type---such as spiral states selected by a continuous ordering wavevector---harmonic spectra vary extensively across configurations and order-by-disorder is effective. In contrast, when configurations are related by local, constraint-preserving rearrangements, their quadratic fluctuation spectra are nearly identical at leading order. Degeneracy lifting then becomes subextensive or higher order in fluctuations, and harmonic selection is parametrically weak.

In canonical three-dimensional spin liquids, the classical degeneracy is not ``accidental'' but enforced by local constraints with a gauge-like structure.
The pyrochlore nearest-neighbor Heisenberg antiferromagnet exemplifies this: the local simplex constraint generates an extensive manifold with algebraic correlations and an absence of conventional ordering at low temperature \cite{MoessnerChalker1998PRL}.
In such settings, harmonic fluctuations frequently fail to provide an extensive selection because the manifold contains continuous families connected by local rearrangements whose fluctuation spectra are nearly identical.
Equivalently, the effective OBD energy is dominated by gauge-like soft modes and loop degrees of freedom, so that degeneracy lifting is subextensive or parametrically weak.
When selection exists, it may occur only at higher order, through anharmonic terms, nonperturbative processes, magnetoelastic couplings, or further-neighbor interactions, and the corresponding energy scale can be dramatically smaller than the microscopic exchange.

Harmonic order-by-disorder can be thermal or quantum. In the thermal case, one compares the fluctuation free energy
\begin{equation}
\Delta F_{\rm th}
=
\frac{T}{2}\sum_{\mathbf k}\ln \det M(\mathbf k),
\end{equation}
where \(M(\mathbf k)\) is the harmonic fluctuation matrix about a given classical configuration. In the quantum case, the corresponding zero-point energy is
\begin{equation}
\Delta E_{\rm q}
=
\frac{1}{2}\sum_{\mathbf k}\omega_{\mathbf k},
\end{equation}
where \(\omega_{\mathbf k}\) are the harmonic fluctuation frequencies. Order-by-disorder is effective when either \(\Delta F_{\rm th}\) or \(\Delta E_{\rm q}\) depends extensively on the position within the classical manifold. In constraint-dominated manifolds, configurations related by local constraint-preserving rearrangements often have very similar harmonic spectra, so the leading selection energy may be small, subextensive, or controlled by additional effects beyond harmonic order. Even when harmonic selection is parametrically weak, real materials may still select order through subleading mechanisms, including anharmonic fluctuations, further-neighbor exchanges, dipolar interactions, or magnetoelastic coupling. The suppression of harmonic order-by-disorder should therefore be viewed as a necessary but not sufficient condition for liquid stabilization.

A particularly sharp realization of this mechanism appears in semiclassical treatments of the quantum pyrochlore antiferromagnet.
Henley showed that harmonic zero-point fluctuations can select a broad subset of collinear states but leave a large residual degeneracy with a gauge-like character: the remaining entropy is subextensive, and the induced effective Hamiltonian is naturally expressed in terms of loop products, i.e., a lattice gauge-theory structure rather than a conventional Landau functional \cite{Henley2006PRL}.
Thus, ‘failure of OBD’ does not mean that fluctuations are irrelevant; rather, they convert the low-energy description from order-parameter selection to constraint-preserving dynamics on a manifold with emergent gauge structure.

\subsection{Dimer and Klein-Point Logic in Three Dimensions}

The above perspective becomes especially sharp in projector-
and dimer-based constructions, where the low-energy sector
is rigorously reduced to a constrained manifold and the
leading dynamics remain internal to that manifold by
construction. In such settings, the usual order-by-disorder
paradigm is largely bypassed: there is no nearby classical
ordered reference state whose harmonic fluctuations must
select among competing configurations. Instead, the
effective problem is posed directly within a constrained
Hilbert space.

The canonical framework is the Rokhsar--Kivelson quantum
dimer model~\cite{Rokhsar-1988}. In these models the
competition between diagonal (``potential'') and off-
diagonal (``kinetic'') terms is explicit: tuning their
relative strength drives transitions between crystalline
phases and liquid regimes, with a special fine-tuned point
at which the ground state becomes an equal-amplitude
superposition within the constrained manifold
\cite{Henley-2004,Moessner-2002}. The conceptual
importance of this construction is
that it makes fully concrete the distinction between merely
suppressing conventional order and actually realizing a
liquid phase supported by internal resonance dynamics.

In three dimensions this logic acquires an especially direct
gauge-theoretic meaning. On bipartite three-dimensional
lattices, quantum dimer models can realize a stable \(U(1)\)
Coulomb phase rather than only an isolated critical point,
with an emergent photon and deconfined monomers
\cite{Moessner-2003,Huse-2003}. On non-bipartite
three-dimensional lattices, by contrast, the corresponding
dimer models can realize gapped \(\mathbb{Z}_2\) liquids.
Thus the gauge structure depends on the bipartiteness of
the underlying dimer problem, just as in two dimensions.

On the pyrochlore lattice, controlled projector and dimer
constructions similarly yield effective Hamiltonians that act
within an extensively degenerate constrained manifold, with
loop- and flippability-based structure rather than conventional
order-parameter energetics
\cite{MoessnerSondhiGoerbig2006PRB}.

In this light, the dimer/Klein-point viewpoint sharpens the conceptual content of ``OBD failure'' in three dimensions:
macroscopic degeneracy is not merely tolerated until fluctuations select an ordered state; instead, the decisive criterion is the existence of a constraint-defined manifold whose dominant dynamics remain internal to that manifold.
In three dimensions, this constraint-preserving dynamics can support a deconfined gauge phase over a finite parameter regime.

This observation suggests a useful diagnostic. If the
classically degenerate manifold is extensive and locally
connected by local constraint-preserving rearrangements,
harmonic order-by-disorder may fail to generate an
extensive Landau selection at leading order. When this
happens, the induced energy differences are subleading
compared with the microscopic exchange scale, and the
dominant low-energy physics can be governed by collective
loop and resonance dynamics internal to the constrained
manifold.

This reframes the role of quantum fluctuations within the
manifold: their leading effect is to generate loop kinetics
and emergent gauge dynamics. The relevant competition is
then between confinement and deconfinement, rather than
among candidate ordered configurations. This naturally leads
to the next stabilizing mechanism, in which spin-orbit
entanglement and multipolar degrees of freedom help engineer
the microscopic kinetic terms needed for such dynamics.


\section{Mechanism III: Microscopic Engineering by Spin--Orbit Entanglement and Multipolar Degrees of Freedom}

\subsection{Dipolar and Multipolar Exchange Physics}

Spin--orbit entanglement reshapes the very notion of a ``spin'' degree of
freedom in 4\(f\)/5\(d\) Mott insulators. In rare-earth magnets, strong
atomic spin--orbit coupling and crystal-field splitting isolate a low-energy
doublet on each site. Let \(\{|\mu\rangle,|\nu\rangle\}\) denote the two
states of this lowest crystal-field doublet. Projecting the physical angular
momentum operator \(J^\alpha\) into the doublet gives
\begin{equation}
\mathcal P J^\alpha \mathcal P
=
\sum_{\mu,\nu\in{\rm doublet}}
\langle \mu | J^\alpha | \nu \rangle
|\mu\rangle \langle \nu| ,
\end{equation}
where \(\mathcal P\) is the projector onto the doublet subspace. This
projected operator can then be represented in terms of effective
pseudospin-\(1/2\) operators.

Because the matrix elements of \(J^\alpha\) within the doublet are strongly
constrained by crystal symmetry, different pseudospin components may
transform as dipoles, quadrupoles, or octupoles. As a result, symmetry
permits highly anisotropic nearest-neighbor interactions---Ising-like
longitudinal terms, transverse pseudospin flips, and bond-dependent
components---even when the underlying exchange is short-ranged. Importantly,
these remain effective pseudospin-\(1/2\) models; the novelty lies not in
enlarging the local Hilbert space, but in how symmetry embeds dipolar and
multipolar operators into the exchange tensor. This structure is explicitly
realized in symmetry-based exchange parameterizations of rare-earth
pyrochlores~\cite{Ross2011PRX,Yan2017PRB}, where longitudinal and transverse
channels arise from distinct doublet matrix elements.

Two classes are especially consequential for three-dimensional quantum liquids.
(i) In \emph{non-Kramers} pyrochlores (e.g.\ Pr$^{3+}$), time-reversal symmetry does not protect the doublet, and transverse components can correspond primarily to \emph{electric quadrupoles} rather than magnetic dipoles.
Microscopically derived pseudospin models then naturally interpolate between an ice-rule sector and quadrupolar or chiral phases, making ``quantum melting'' of an ice manifold a generic possibility rather than a fine-tuned accident \cite{OnodaTanaka2010PRL,OnodaTanaka2011PRB}.
(ii) In \emph{dipole--octupole} (DO) Kramers doublets (realized for certain $4f$ ions), two pseudospin components transform as magnetic dipoles while the third transforms as a component of an octupole tensor.
This representation changes which interactions act as ``transverse'' dynamics of an ice-like constraint and, crucially, which excitations couple to conventional probes \cite{HuangChenHermele2014PRL}.

A useful way to view spin--orbit entanglement is therefore as an
interaction-engineering principle, rather than as a guarantee of liquid
behavior. Its role is to make additional anisotropic and multipolar exchange
channels symmetry-allowed. When the dominant exchange hierarchy has already
established a constrained manifold, these channels can provide the
microscopic kinetic terms needed for constraint-preserving quantum dynamics.
When they instead favor competing order, the same couplings can destabilize
the liquid regime. Thus spin--orbit entanglement can help implement the
constraint-first mechanism, but only when the resulting anisotropic exchange
acts predominantly within the constrained sector rather than selecting a
symmetry-breaking state~\cite{GingrasMcClarty2014RoPP,RauGingras2019ARCMP}.

\subsection{Multipolar Constraints and Novel Quantum Liquids}

Multipolar degrees of freedom do more than renormalize parameters of familiar spin models: they can change the \emph{structure} of the low-energy Hilbert space and the nature of the emergent gauge theory.
In non-Kramers systems, the ``transverse'' terms are often quadrupolar operators.
Within an ice manifold, such terms generate tunneling processes that need not correspond to simple dipole flips; instead they connect constraint-satisfying configurations through multipolar pathways and can drive transitions to quadrupolar-ordered states or stabilize liquid regimes whose natural description involves gauge fields coupled to multipolar matter \cite{OnodaTanaka2010PRL,OnodaTanaka2011PRB}.
In DO-doublet systems, the octupolar component can act as the transverse field of an ice constraint while remaining partially ``hidden'' to dipolar probes, yielding \emph{symmetry-enriched} U(1) quantum spin liquids where the emergent electric and magnetic sectors transform nontrivially under crystal symmetries \cite{HuangChenHermele2014PRL}.

This physics becomes experimentally sharp when dipolar scattering signatures of an underlying gauge structure coexist with dynamics set by multipolar terms.
In Ce$_2$Zr$_2$O$_7$, inelastic neutron scattering measurements report a continuum of magnetic excitations consistent with quantum spin-ice-like dynamics, in a setting where the microscopic doublet structure and anisotropic exchange are central, and the interpretation connects naturally to DO-based U(1) gauge theories rather than to a purely dipolar Ising model \cite{Gaudet2019PRL}. Symmetry-based parameterizations of anisotropic exchange on the pyrochlore lattice emphasize that the allowed coupling space is high-dimensional and hosts many competing ordered and liquid regimes. Empirically, several leading candidate materials appear to sit near phase boundaries where constraint manifolds, multipolar tunneling, and proximate ordering tendencies compete, which can be advantageous for stabilizing constraint-preserving kinetics while suppressing premature selection~\cite{Ross2011PRX,Yan2017PRB,RauGingras2019ARCMP,Gresista-2025}. More generally, the neighboring ordered states of pyrochlore quantum spin ice need not be viewed as unrelated competitors, but can arise as confinement transitions of the U(1) spin liquid itself, a perspective developed explicitly in monopole-condensation approaches to Pr$_2$Ir$_2$O$_7$ and Yb$_2$Ti$_2$O$_7$ \cite{Chen-2016}. From the mechanism perspective, this proximity is not a nuisance but an opportunity: it is precisely where multipolar dynamics can be strong enough to generate coherent gauge-field kinetics while remaining constraint-respecting at leading order.

Spin–orbit entanglement therefore offers a route to three-dimensional quantum liquids that is qualitatively distinct from ``more frustration'' or ``smaller $S$''.
By encoding dipolar and multipolar operators into the effective pseudospin algebra, it simultaneously creates classical constraints and supplies the symmetry-allowed quantum dynamics that act within them.
In three dimensions, the resulting compact gauge structures---whether conventional U(1) spin ice or symmetry-enriched variants tied to multipolar representations---can remain deconfined over extended regimes, making multipolar pyrochlores a concrete materials setting for realizing and diagnosing genuine three-dimensional quantum spin liquids~\cite{SavaryBalents2012PRL,SavaryBalents2016RoPP}.

A working criterion is that the crystal-field doublet enforces that transverse operators transform as higher multipoles under the lattice symmetry, constraint-preserving tunneling processes can be enhanced without necessarily amplifying conventional dipolar ordering channels. In such circumstances, gauge kinetics generated by multipolar exchange can be of order comparable to the dominant Ising scale, whereas symmetry-allowed dipolar ordering terms appear only at subleading order.
The effective low-energy theory is therefore naturally driven toward a regime where coherent constraint-preserving tunneling competes with, or overwhelms, conventional symmetry breaking. Whether this separation is realized in practice depends sensitively on the allowed exchange tensor and symmetry-allowed invariants; dipolar instabilities may still appear at comparable order in less restrictive settings.

A controlled separation between constraint enforcement and multipolar
kinetics arises most transparently in rare-earth pyrochlores where
crystal-field splitting isolates a well-defined doublet separated from
excited states by an energy $\Delta_{\mathrm{CEF}} \gg J$.
Projection of the microscopic superexchange into this doublet generates
an anisotropic exchange tensor whose leading longitudinal component
(e.g.\ $J_{zz}$ in Eq.~\eqref{eqn:general_ham} enforces the ice or simplex constraint at
order $J$, while transverse multipolar terms originate from off-diagonal
matrix elements within the doublet and scale parametrically as
$J_{\perp} \sim J \times (\lambda_{\mathrm{mix}})$,
where $\lambda_{\mathrm{mix}}$ encodes the symmetry-allowed admixture of
multipolar character. In favorable representations—such as non-Kramers
or dipole–octupole doublets—the longitudinal component can dominate at
leading order while transverse multipolar processes remain symmetry-allowed
but do not directly generate dipolar ordering channels.
In this regime the hierarchy
$J_{zz} \gtrsim J_{\perp} \gg J_{\mathrm{dipolar\ instability}}$
is microscopically controlled by crystal-field structure and symmetry,
rather than by fine tuning of unrelated couplings.

The implication is not merely quantitative. More fundamentally, spin--orbit entanglement alters the \emph{symmetry representation content} of the local doublet, i.e.\ which combinations of dipolar/quadrupolar/octupolar operators act as the effective pseudospin components and how they transform under the space group and time reversal. The resulting deconfined phase may carry symmetry-enriched quantum numbers tied to multipolar representations, altering both the spectrum of excitations and their experimental visibility. Three dimensions then provide the final stabilizing ingredient: compact gauge fields generated by such multipolar kinetics admit stable Coulomb phases, so that strong transverse dynamics need not trigger confinement.

\section{Design Patterns for Stabilizing 3D Quantum Spin Liquids}

The preceding sections can be recast as design criteria.
Here, we reformulate them as concrete design criteria tied directly to lattice geometry and exchange hierarchy.
The central question is operational: given a three-dimensional magnetic network and a symmetry-allowed exchange tensor, under what structural conditions is a liquid regime parametrically favored over symmetry breaking?

\subsection{Network Topology and Loop Dynamics}

Constraint-dominated liquids in three dimensions almost invariably arise on lattices built from corner-sharing simplices.
The pyrochlore lattice consists of corner-sharing tetrahedra; the hyperkagome lattice of corner-sharing triangles embedded in three dimensions \cite{Okamoto2007PRL,Lawler2008PRL}.
In both cases, the dominant nearest-neighbor exchange enforces a local simplex constraint (e.g.\ $\sum_{i\in \text{tet}} \mathbf{S}_i = 0$ for Heisenberg pyrochlore or an ice-rule constraint for Ising anisotropy).
The relevant geometric variable is not frustration alone, but the density and connectivity of independent closed loops that preserve the constraint.

On the pyrochlore lattice, minimal loops involve six-site hexagons. An extensive set of such elementary hexagonal loops exists throughout the lattice, providing the natural support for ring-exchange processes within the constrained manifold.
Quantum processes generated at higher order in perturbation theory act precisely around such loops, producing effective ring-exchange terms whose number scales with system volume \cite{HermeleFisherBalents2004PRB}. Such loop dynamics can be represented schematically by operators of the form
\begin{equation}
\mathcal O_{\ell}
= S_1^{+} S_2^{-} S_3^{+} S_4^{-} S_5^{+} S_6^{-}
+ \text{h.c.},
\end{equation}
acting around elementary hexagons $\ell$. These operators flip spins coherently along a closed loop while preserving the local simplex constraint at leading order. Because these loops intersect and tile three-dimensional space, loop kinetics generate a macroscopic gauge field rather than isolated resonances.
The same principle governs hyperkagome networks, where loop connectivity frustrates simple ordering wavevectors and favors extended resonance patterns \cite{Lawler2008PRL,Zhou2008PRL}.

Recent materials studies already indicate that this design axis extends beyond the canonical pyrochlore and hyperkagome settings. In PbCuTe$_2$O$_6$, nearly balanced interactions generate the hyper-hyperkagome network, a denser three-dimensional corner-sharing-triangle architecture in which each site belongs to three triangles and the shortest nontrivial loops are reduced to four- and six-site circuits \cite{Chillal2020NatComm}. In K$_2$Ni$_2$(SO$_4$)$_3$, the relevant magnetic network can be viewed as tetrahedra arranged on a trillium backbone, revealing a distinct simplex-based route to strong three-dimensional dynamics and proximity to a liquid regime \cite{Gonzalez2024NatComm}. These examples suggest that the decisive geometric variable is not allegiance to a canonical lattice family, but how a given simplex network reshapes local constraint connectivity and the available manifold of loop or resonance processes.

Thus the geometric stabilizer is the existence of an extensive network of constraint-preserving loops that (i) preserves the leading local constraint and (ii) generates non-commuting resonance processes whose number scales as $\mathcal{O}(N)$.
When these conditions hold, the effective Hamiltonian is naturally expressed in terms of loop operators, and the low-energy description maps onto a lattice gauge theory rather than a Landau functional.

\subsection{Liquid Islands in Coupling Space}

Topology alone is insufficient; the exchange hierarchy must position the
system near a classically degenerate manifold. In anisotropic rare-earth
pyrochlores, the symmetry-allowed nearest-neighbor Hamiltonian takes the
form~\cite{Ross2011PRX}
\begin{equation}
\label{eqn:general_ham}
\begin{aligned}
\mathcal{H} &= \sum_{\langle ij\rangle} \Big[
 J_{zz} S_i^z S_j^z
- J_{\pm}(S_i^+ S_j^- + \mathrm{h.c.}) \\
&\quad + J_{\pm\pm} (\gamma_{ij} S_i^+ S_j^+ + \mathrm{h.c.})
+ J_{z\pm} \big( S_i^z \zeta_{ij} S_j^+ + \mathrm{h.c.} \big)
\Big] .
\end{aligned}
\end{equation}
Here the spin components are defined in their local pyrochlore frames, and
\(\gamma_{ij}\) and \(\zeta_{ij}\) are bond-dependent phase factors fixed by
lattice symmetry. When \(J_{zz}\) dominates, the classical manifold is
ice-like. The transverse terms \(J_{\pm}\), \(J_{\pm\pm}\), and \(J_{z\pm}\)
then generate quantum dynamics and effective interactions within, or
virtually connecting, this manifold. Phase diagrams of this model show that
quantum liquid regimes typically appear near boundaries between multiple
ordered states, where classical ground-state energies are nearly degenerate
\cite{Yan2017PRB}.

The essential microscopic requirement is a hierarchy of scales,
\begin{equation}
J_{zz} \gg |J_{\pm}|, |J_{\pm\pm}|, |J_{z\pm}| ,
\end{equation}
together with the more restrictive condition that the leading
constraint-preserving kinetic scale not be overwhelmed by lower-order
ordering potentials generated within the constrained manifold. For quantum
spin ice, the elementary hexagonal ring exchange generated by \(J_{\pm}\)
has the scale
\begin{equation}
K \sim \frac{J_{\pm}^3}{J_{zz}^2},
\end{equation}
up to a numerical prefactor. By contrast, the \(J_{z\pm}\) coupling can
generate diagonal interactions already at second order in perturbation
theory, including effective further-neighbor Ising terms of scale
\begin{equation}
V_{z\pm}\sim \frac{J_{z\pm}^2}{J_{zz}} ,
\end{equation}
which favor conventional ordering. A necessary condition for a robust
quantum spin-ice regime is therefore that the loop kinetic scale remain
competitive with such ordering tendencies, for example
\begin{equation}
K \gtrsim V_{z\pm},
\qquad
\frac{J_{\pm}^3}{J_{zz}^2}
\gtrsim
\frac{J_{z\pm}^2}{J_{zz}},
\end{equation}
up to lattice-dependent numerical factors and additional higher-order
cross-terms involving \(J_{\pm\pm}\).

In the gauge-theory description, the relevant energetic separation is not
that the magnetic monopole gap exceeds the gauge-charge gap or a Higgs
scale. Rather, the constraint-violating electric charges must remain costly
on the scale of the ring dynamics. In quantum spin ice, electric charges
associated with ice-rule violations are set by the constraint scale
\(J_{zz}\), whereas the magnetic monopole gap and photon bandwidth are
controlled by the much smaller ring-exchange scale
\(K\sim J_{\pm}^3/J_{zz}^2\). A stable Coulomb regime requires the system to
remain on the weak-coupling side of the compact \(U(1)\) gauge theory, with
neither electric charges nor magnetic monopoles proliferating.

A crucial refinement concerns disorder and transverse-field inhomogeneity.
The relevant scale governing coherent gauge dynamics is not the dominant
Ising exchange \(J_{zz}\), but the much smaller ring-exchange amplitude
\(K\). Random transverse fields or exchange disorder can generate lower-order
potential terms that act directly within the constrained manifold. Because
such terms may appear at lower order than \(K\), they can dominate the
effective dynamics even when their bare strength is small compared with
\(J_{zz}\). The criterion for a clean Coulomb phase is therefore not merely
that the disorder scale be small compared with \(J_{zz}\), but that it remain
small compared with the intrinsic gauge-kinetic scale \(K\)~\cite{Benton-2018}.

Disorder and additional weak interactions have two distinct roles. First,
they can favor conventional ordering or freezing, thereby truncating the
liquid regime. Second, when sufficiently weak, they may suppress a nearby
ordered instability without destroying the constraint structure. The latter
case can enlarge an apparent liquid-like window, but it should not be
confused with intrinsic deconfinement. Establishing a genuine QSL therefore
requires evidence that the low-energy response is governed by fractionalized
excitations and emergent gauge dynamics, rather than by disorder-broadened
short-range correlations.

Stability of the liquid phase therefore requires a finite window in which
constraint-preserving kinetics dominate over ordering potentials, disorder,
and confinement tendencies. In this window, harmonic order-by-disorder
selection is weak because several competing instabilities have nearly
similar quadratic fluctuation spectra, while the leading low-energy dynamics
remain internal to the constrained manifold. The resulting regime in
parameter space is not a fine-tuned point but a finite domain bounded by
ordered, frozen, or confined phases---a ``liquid island''.

In gauge-theory language, such an island corresponds to a Coulomb phase with
finite emergent gauge stiffness and gapped magnetic monopole excitations.
Here the gauge stiffness refers to the coefficient of the Maxwell term in the
low-energy compact \(U(1)\) gauge theory; together with the conjugate
electric-field stiffness it controls the photon velocity, but it should not
be identified with the monopole gap. The Coulomb regime can be lost through
monopole proliferation, condensation of gauge-charged matter, or perturbations
that explicitly break the emergent gauge description. When matter fields
carrying fundamental gauge charge are present, the distinction between Higgs
and confined regimes need not define a sharp phase boundary, in accordance
with Fradkin--Shenker continuity. For the present purposes, the important
point is that all of these routes destroy the deconfined Coulomb description.

For compact \(U(1)\) gauge structure, three dimensions enlarge the possible
liquid domain relative to two-dimensional analogues because, once
loop-generated gauge dynamics dominate, the Coulomb phase is stable at weak
coupling rather than generically confined.

These requirements can be stated as a hierarchy of requirements. A three-dimensional lattice favors liquid stabilization when:
(i) its minimal constraint-preserving loops form an extensive set whose
number scales with system size; (ii) the dominant exchange enforces a local
constraint at a large energy scale \(J_{\rm constraint}\); and (iii)
symmetry-allowed transverse processes generate loop kinetics of scale \(K\)
that remain finite in the thermodynamic limit and exceed the leading
ordering or disorder-induced potentials within the constrained manifold.
When, in addition, competing classical instabilities are nearly degenerate,
harmonic selection can become parametrically weak and the effective
low-energy theory can resolve into a deconfined gauge sector. Three-dimensional
geometry then stabilizes the liquid by permitting this gauge sector to remain
in a Coulomb phase rather than confining.

Putting these elements together suggests a hierarchy governing stabilization.
Geometry defines the constraint manifold; symmetry determines which
transverse processes act within it; harmonic selection must remain weak on
the scale of the intrinsic gauge kinetics; disorder and residual interactions
must not overwhelm the ring dynamics; and dimensionality stabilizes the
resulting gauge sector. The mechanisms discussed above differ in microscopic
origin but converge on this structural logic.

\section{Diagnostics and Falsification in Three Dimensions}

A three-dimensional quantum spin liquid (QSL) is not defined experimentally by the absence of Bragg peaks. It is instead characterized by a low-energy sector that cannot be consistently captured within a description based on (possibly broadened) magnons around a nearby symmetry-breaking state. In practice, this becomes a falsification program: one must exclude (i) static order below instrumental resolution, (ii) glassy freezing, and (iii) disorder-dominated paramagnetism, while establishing signatures consistent with deconfinement and an emergent gauge structure~\cite{Zhou2017RMP,Wen2019npjQM,SavaryBalents2016RoPP}.

While quantum spin ice (and U(1) Coulomb phases) provide the cleanest 3D case study because they support an emergent photon, the broader diagnostic logic also applies to $\mathbb{Z}_2$ spin liquids. In 3D $\mathbb{Z}_2$ QSLs, gauge flux excitations are loop-like, and the corresponding lattice gauge theories admit finite-temperature confinement transitions associated with flux-loop proliferation; experimentally, one therefore focuses less on a photon mode and more on (i) continua that cannot be consistently reconstructed from multi-magnon or disorder-broadened quasiparticle processes, (ii) symmetry/selection rules in dynamical structure factors and Raman response, and (iii) field/pressure evolution that is consistent with deconfinement versus confinement/Higgs scenarios rather than with broadened magnons alone~\cite{Nasu-2014,Smith-2015,Perreault-2015}.

The most treacherous mimic is a constrained classical spin liquid with slow dynamics and weak disorder, which can reproduce diffuse scattering and persistent spin dynamics over extended temperature windows. Importantly, because the relevant gauge-kinetic scale in quantum spin ice is the ring-exchange amplitude $K \sim J_{\pm}^3/J_{zz}^2$, even weak transverse-field disorder can overwhelm coherent tunneling if its magnitude exceeds $K$, a hierarchy emphasized in theoretical studies of disordered quantum spin ice \cite{Benton-2018}. Establishing quantum coherence therefore requires cross-consistency between thermodynamics, spectroscopy, and field-tuned responses within the same sample.

\subsection{Fractionalization versus Disorder}

The basic distinction is that disorder removes order by randomizing local environments, whereas a QSL removes order by endowing the low-energy Hilbert space with topological or gauge structure. This distinction is naturally phrased in the language of topological order and emergent gauge fields~\cite{Wen-1990,Wen-2002,Sachdev-2019}. Consequently, the most discriminating diagnostics are those that probe \emph{structure} (constraints, gauge modes, symmetry fractionalization), not merely the lack of static moments.

For three-dimensional Coulombic liquids (notably quantum spin ice), the key structural fingerprint is a Gauss-law constraint that survives into the quantum regime.
In reciprocal space this appears as pinch-point correlations in the equal-time structure factor; in a quantum-coherent regime these pinch points are suppressed and rounded in a manner consistent with lattice gauge theory predictions, rather than being destroyed arbitrarily as would typically occur in simple disorder scenarios that violate the underlying constraint. Disorder that preserves the divergence-free constraint can, however, broaden pinch points in a more controlled manner with distinguishable signatures~\cite{GingrasMcClarty2014RoPP,Benton2012PRB,Sibille2018NatPhys}.
This is a falsifiable statement: the location, anisotropy, and line shape are constrained by the divergence-free condition and have limited flexibility within the constraint-consistent description.

Experimental indications of such quantum Coulomb physics have been reported in Pr$_2$Zr$_2$O$_7$, where inelastic neutron scattering reveals pinch-point-like correlations coexisting with low-energy dynamical spectral weight and activated thermodynamic behavior consistent with quantum spin-ice–like dynamics~\cite{Kimura-2013}. More recently, $\mathrm{Ce_2Zr_2O_7}$ has emerged as a leading dipole–octupole quantum spin-ice candidate.  Inelastic neutron scattering reports low-energy continua and constraint-consistent correlations compatible with a proximate U(1) quantum spin liquid regime. Diffuse-scattering and spectroscopic studies emphasize the role of dipole–octupole doublets in generating transverse dynamics within an ice manifold, sharpening the connection between multipolar exchange and emergent gauge structure \cite{CeZr-2024,CeZr-PRX-2025}. These observations suggest that the classical Coulomb manifold can persist into a regime where quantum fluctuations generate coherent dynamics rather than static freezing.

Fractionalization requires deconfined excitations. In
U(1) spin ice these include gapped electric charges, gapped
magnetic monopoles, and a gapless photon mode~\cite{HermeleFisherBalents2004PRB,SavaryBalents2012PRL}. Here, it
is essential to be conservative about continua: broad
inelastic scattering is not, by itself, evidence for
fractionalization, because multi-magnon processes,
proximity to criticality, or disorder-broadened modes can
all generate continua over limited windows.
What is harder to mimic is a \emph{coherent} near-zero-energy linearly dispersing mode with the polarization and momentum dependence expected for an emergent photon, together with independent evidence that the system remains in a constraint-governed regime \cite{GingrasMcClarty2014RoPP,Benton2012PRB}. A complementary target is the spectroscopy of magnetic monopoles themselves: in pyrochlore U(1) spin liquids,
their continuum structure has a distinct kinematic and
symmetry organization that can, in principle, supplement
photon-based diagnostics \cite{Chen-2017}. At long wavelengths, this emergent photon exhibits a linear dispersion
\begin{equation}
\omega(\mathbf k) = c\, |\mathbf k| ,
\end{equation}
with photon velocity $c$ determined by the ratio of electric and magnetic stiffnesses in the effective Maxwell action. The spectral weight is concentrated in the transverse channel, reflecting the divergence-free constraint. This linear mode is distinct from magnon Goldstone modes, as it arises from gauge redundancy rather than spontaneous symmetry breaking. Crucially, quenched disorder can broaden or renormalize
constraint correlations, but it does not generate a
collective transverse gauge mode with a well-defined
Maxwell stiffness. The emergence of a coherent
linearly dispersing mode with constrained polarization
structure therefore signals collective gauge dynamics
rather than static randomness.

A practical caveat is that in highly frustrated magnets with extremely weak selection (e.g.\ parametrically small order-by-disorder scales), an almost-gapless pseudo-Goldstone mode from a nearby ordered state could, over an intermediate window, mimic an approximately linear dispersion. Disentangling these scenarios requires using \emph{structure} rather than dispersion alone: the emergent photon has a characteristic transverse polarization structure tied to the Gauss-law constraint and exhibits a distinctive evolution of pinch-point line shapes and spectral weight under tuning (field/pressure) that differs from conventional pseudo-Goldstone physics~\cite{Benton2012PRB,Fradkin-2013}.

\subsection{Experimental Probes}

\textit{Neutron scattering} is the main probe of the constraint sector and the excitation spectrum. A credible 3D-QSL claim based on neutrons should satisfy three increasingly stringent checks:

(i) \emph{Elastic channel (conservative criterion):} no resolution-limited magnetic Bragg peaks and no growth of quasi-static diffuse scattering indicative of freezing.
(Strictly speaking, weak symmetry breaking can in principle coexist with fractionalization in symmetry-enriched or Higgs-descended phases; nevertheless, the absence of Bragg peaks remains the cleanest and most conservative empirical filter for identifying fully disordered 3D-QSL regimes.)

(ii) \emph{Quasi-elastic / equal-time correlations:} reciprocal-space patterns consistent with a constraint manifold (pinch points or their quantum-suppressed descendants), including correct symmetry locations and anisotropic line shapes \cite{Benton2012PRB,Sibille2018NatPhys}.

(iii) \emph{Inelastic channel:} dynamics that are incompatible with a sum of broadened magnons.
For quantum spin ice, this means looking for the predicted near-zero-energy photon response (small velocity, weak spectral weight) and its polarization structure; polarized neutron analysis is particularly valuable because it separates magnetic from nuclear/incoherent contributions and constrains the photon-like channel more sharply \cite{Gao2025NatPhys,Gaudet2019PRL}. 

Beyond linear response, multidimensional and nonlinear spectroscopic
probes have been proposed as sharper diagnostics of fractionalization.
Higher-order response functions can access symmetry channels and
correlation structures that are invisible to conventional two-point
structure factors, and may help distinguish intrinsic fractionalized
continua from multi-magnon or disorder-broadened backgrounds.
For example, THz 2D coherent spectroscopy has been shown to resolve
fractional excitation continua in spin systems by exploiting spinon
echo techniques \cite{Wan-2019}, and microscopic analyses
demonstrate how nonlinear spectral features can distinguish different
spin-liquid scenarios \cite{Sim-2023,Ginley-2024}. Such tools remain experimentally
challenging in three-dimensional magnets, but they provide promising
future directions for sharpening the identification of deconfined
gauge structure.

\textit{Thermodynamics} constrains the density of low-energy states.
In a U(1) Coulomb phase, the photon gives a $C \sim T^3$ contribution at sufficiently low temperature, but this is only meaningful when nuclear Schottky and disorder contributions are under control and when the same sample shows constraint-consistent correlations \cite{GingrasMcClarty2014RoPP}.
Field dependence is often more diagnostic than zero-field scalings: fields couple to gauge charges and can induce crossovers or confinement/Higgs-like transitions, which should reshape both diffuse scattering and low-energy spectral weight in correlated ways. Within the gauge description, a confinement or Higgs transition corresponds to condensation of a matter field $\Phi$ coupled minimally to the emergent gauge
field $A_\mu$,
\begin{equation}
\mathcal L_{\mathrm{Higgs}} =
\left|(\partial_\mu - i A_\mu)\Phi\right|^2
+ r |\Phi|^2 + u |\Phi|^4 .
\end{equation}
Tuning $r$ across zero drives the Higgs transition, reshaping both the diffuse structure factor and the low-energy spectral weight in a manner tightly constrained by gauge invariance.

\textit{Local probes} ($\mu$SR, NMR) are powerful at \emph{excluding} static order and glassiness, but are not intrinsically specific to fractionalization.
A long-time muon polarization plateau or persistent relaxation can indicate dynamic magnetism, yet similar phenomenology can occur in disorder-dominated regimes.
They are best used as falsification filters: ruling out freezing and placing upper bounds on static internal fields, ideally on the same crystals used for neutron measurements \cite{Wen2019npjQM}.

\subsection{Limits of Current Diagnostics}

No single measurement can establish a three-dimensional QSL.
Each major observable has a known loophole:
pinch-point-like diffuse scattering can persist in constrained \emph{classical} regimes above a quantum scale,
continua can arise without fractionalization,
and power laws can be accidental over finite windows.
Even the emergent photon, the closest analogue of a ``smoking gun'' for U(1) spin ice, is experimentally challenging because its spectral weight is weak and it lives close to zero energy where backgrounds and resolution effects are most severe \cite{GingrasMcClarty2014RoPP,Gao2025NatPhys}.

The most robust standard is therefore \emph{cross-consistency under tuning}.
A credible identification requires that (a) constraint-consistent equal-time correlations,
(b) excitation spectra compatible with deconfinement (and not with broadened magnons),
and (c) systematic evolution under field/pressure/chemical tuning,
all cohere into a single effective description.
In three dimensions, stabilization is often easier than detection: gauge structure can be robust while its experimental signatures remain subtle. For this reason, a conservative diagnostic posture is essential; without it, claims of a 3D quantum spin liquid remain incomplete.


\section{Outlook: Emerging Opportunities}

The preceding sections point to a specific conclusion: three-dimensional quantum
spin liquids can occur when a local constraint manifold is connected by
symmetry-allowed loop kinetics before ordinary ordering tendencies take over.
The next step is to test this sequence in new symmetry classes, lattice
geometries, and gauge structures.

\subsection{Materials Considerations}

For materials work, the sequence is straightforward.
First-principles electronic-structure calculations constrain the exchange tensor and crystal-field scheme.
Projection into the lowest doublet yields an effective pseudospin model whose anisotropy structure can be classified by symmetry. The central materials question is then not whether frustration exists, but whether the dominant exchange enforces a local constraint at leading order and whether symmetry permits transverse or multipolar processes that act within that manifold. In practice, this separation is rarely clean in real compounds, but the hierarchy remains a useful guide. A further practical refinement concerns the relevant energy scale
that must dominate over imperfections.
In quantum spin-ice regimes, coherent gauge dynamics are governed
not by the dominant Ising exchange $J_{zz}$ itself,
but by the ring-exchange scale
$K \sim J_{\pm}^3/J_{zz}^2$ generated within the constrained manifold.
Material design must therefore satisfy the hierarchy
\begin{equation}
\text{(disorder scale)} \ll K,
\end{equation}
rather than merely $\text{(disorder scale)} \ll J_{zz}$.
Because $K$ is parametrically smaller than $J_{zz}$,
this imposes substantially more stringent requirements on crystal
homogeneity and transverse-field uniformity than a naive exchange-scale
comparison would suggest \cite{Benton-2018}.

This perspective has already guided the identification of quantum spin-ice candidates and dipole--octupole pyrochlores, but its scope is broader.
Breathing pyrochlores, where alternating tetrahedra carry distinct exchange scales, offer tunable proximity to constraint manifolds and may host variants of Coulombic phases in regimes where classical degeneracy is only partially lifted.
More generally, symmetry-enriched U(1) liquids can arise when spin--orbit entanglement ties gauge charges to crystal symmetry representations, modifying their transformation properties and experimental signatures.
Lattice symmetry can therefore be used as a resource for engineering gauge structure.

A more speculative extension concerns three-dimensional chiral quantum spin
liquids. Recent work on the pyrochlore lattice suggests that further-neighbor
exchange, ring-exchange processes, spin-anisotropic couplings, or explicit
chiral interactions can support such phases~\cite{Lapa-2012,Gomez-2024,Kranitz-2026,Iqbal-2019,Iqbal-2017}.
From a materials perspective, the relevant target is not chirality in the
crystallographic sense alone, but a regime with noncoplanar magnetic tendencies
and finite scalar spin chirality at intermediate scales, so that quantum
fluctuations can melt dipolar order without immediately destroying the chiral
structure.

\subsection{New Lattice Architectures}

\begin{figure}[t]
    \centering

    \begin{minipage}[c][0.28\textwidth][c]{0.32\textwidth}
        \centering
        \begin{overpic}[width=\linewidth]{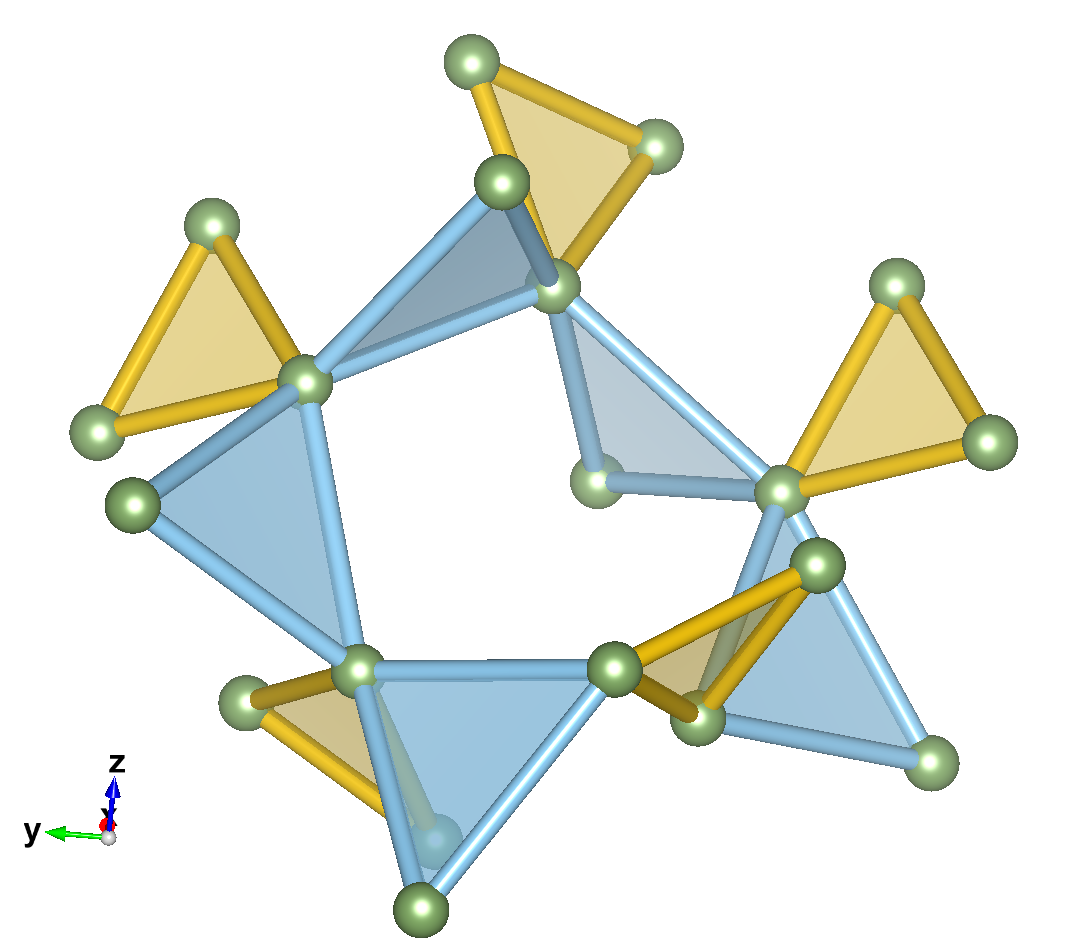}
            \put(4,94){\bfseries (a)}
        \end{overpic}
    \end{minipage}\hfill
    \begin{minipage}[c][0.28\textwidth][c]{0.32\textwidth}
        \centering
        \begin{overpic}[width=\linewidth]{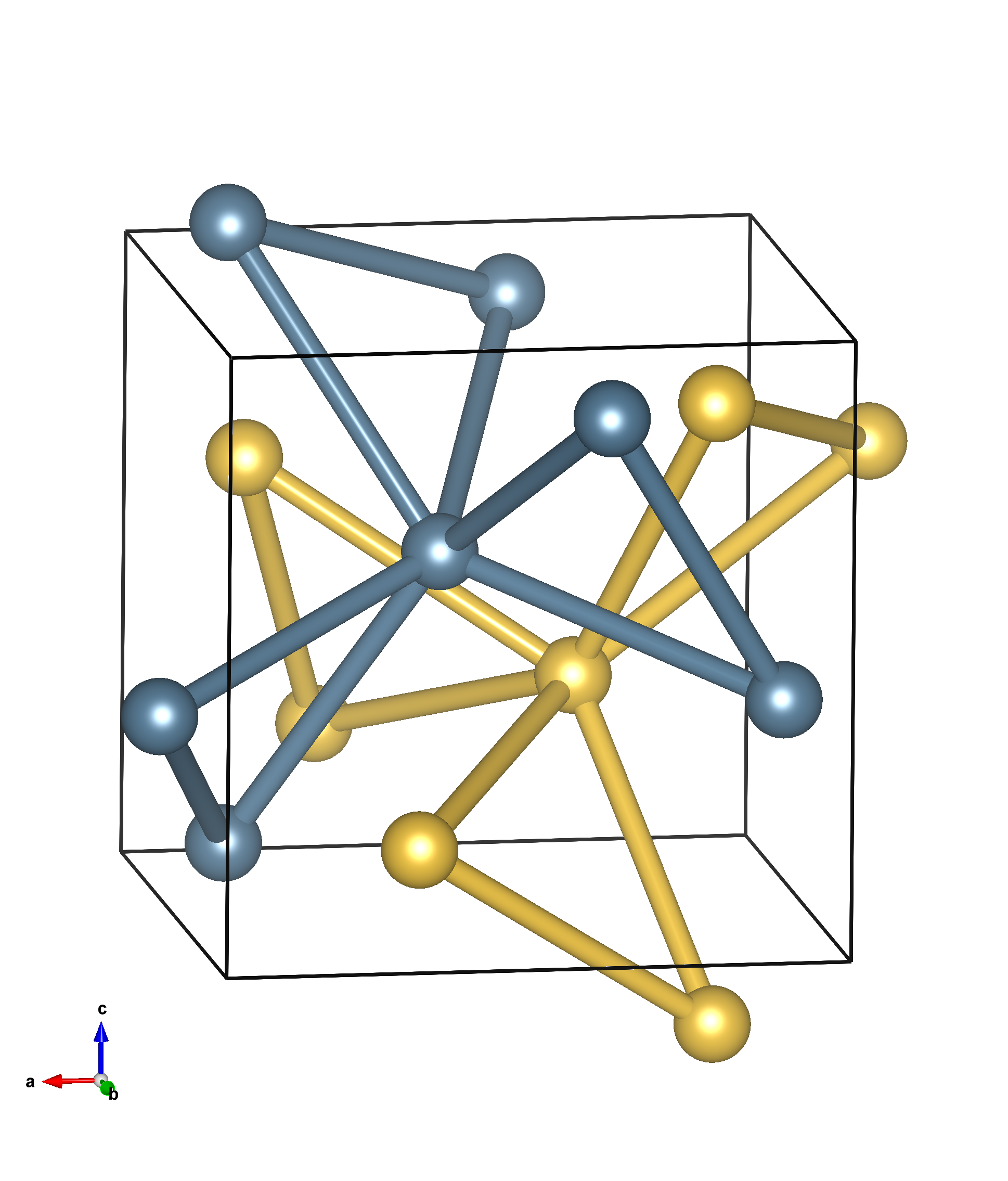}
            \put(4,94){\bfseries (b)}
        \end{overpic}
    \end{minipage}\hfill
    \begin{minipage}[c][0.28\textwidth][c]{0.32\textwidth}
        \centering
        \begin{overpic}[width=\linewidth]{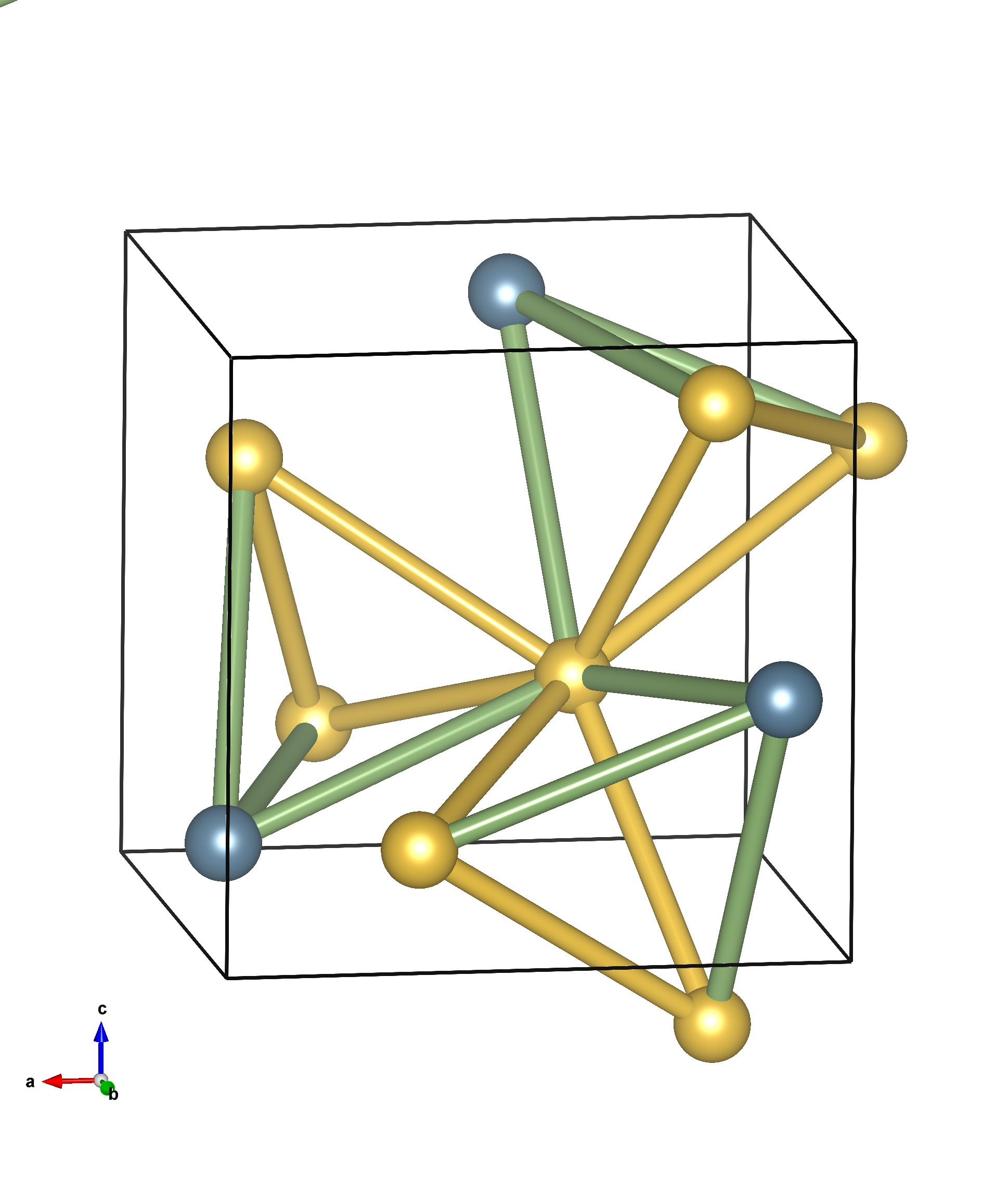}
            \put(4,94){\bfseries (c)}
        \end{overpic}
    \end{minipage}

    \caption{Examples of noncanonical three-dimensional frustrated architectures discussed in the text. 
    (a) Hyper-hyperkagome magnetic structure relevant to PbCuTe$_2$O$_6$, drawn from the crystallographically equivalent Cu$^{2+}$ sites with the dominant first- and second-neighbor interactions. One set of bonds forms isolated triangles, while the other generates the connected three-dimensional network of corner-sharing triangles. 
    (b) Two interpenetrating trillium sublattices relevant to K$_2$Ni$_2$(SO$_4$)$_3$, shown in different colors for clarity. 
    (c) Inter-sublattice connectivity generates tetrahedral units arranged on a trillium backbone, giving rise to a tetra-trillium lattice. 
    Together, these examples illustrate how moving beyond canonical pyrochlore and hyperkagome settings can reshape the local constraint structure in three dimensions through distinct simplex geometries. Panel (a) is adapted from Ref.~\cite{Chillal2020NatComm}; panels (b,c) are adapted from Ref.~\cite{Gonzalez2024NatComm}.}
    \label{fig:new_3d_architectures}
\end{figure}

Beyond canonical pyrochlores and hyperkagome networks, the next advances may come from expanding the geometric library of three-dimensional frustrated magnets rather than revisiting familiar platforms alone. The relevant question is not simply whether a lattice is frustrated, but what kind of local constraint algebra and loop dynamics its simplex structure permits. In this regard, several recently explored architectures suggest that the design space of three-dimensional quantum spin liquids is already broader than the standard pyrochlore-centered narrative.

A first example is the hyper-hyperkagome network realized in PbCuTe$_2$O$_6$, where nearly balanced interactions generate a denser three-dimensional lattice of corner-sharing triangles than in the hyperkagome case (Fig.~\ref{fig:new_3d_architectures} (a)). Each site participates in three triangles rather than two, and the shortest nontrivial loops are reduced to four- and six-site circuits~\cite{Chillal2020NatComm}. This changes both the local constraint connectivity and the natural space of resonance processes, suggesting that three-dimensional triangle-based liquids may not be exhausted by the hyperkagome paradigm.

A second direction is provided by langbeinite materials such as K$_2$Ni$_2$(SO$_4$)$_3$, where the relevant magnetic network can be viewed as tetrahedra arranged on a trillium backbone, yielding a tetra-trillium architecture (Fig.~\ref{fig:new_3d_architectures}(b) and (c)) \cite{Gonzalez2024NatComm}. The key lesson here is not necessarily the realization of a fully developed spin liquid in the material itself, but the emergence of a proximate liquid regime associated with a tetra-trillium architecture~\cite{Gonzalez-2025}. This points to a broader family in which tetrahedral and triangular simplexes are recombined into new three-dimensional tilings whose constraint manifolds and loop statistics differ qualitatively from those of pyrochlores.

Another promising example is the octochlore lattice, a three-dimensional network of corner-sharing octahedra. In contrast to pyrochlore spin ice, where the canonical local object is the tetrahedron, the octochlore geometry reorganizes the constraint structure around octahedral units and thereby enlarges the space of possible Coulomb phases. Recent work has shown that such systems can support unusual Coulomb spin liquids, including regimes with enhanced pinch-point structure and higher-rank-like constraint phenomenology~\cite{Benton-2021}. Very recent studies further suggest that Ising-like octochlore magnets may provide a route to fracton-like classical spin liquids with restricted-mobility excitations, reinforcing the broader lesson that expanding the simplex geometry can qualitatively enlarge the space of emergent gauge structures~\cite{Stern2026}.

More generally, recent theoretical work on trillium-derived line-graph lattices indicates that such noncanonical three-dimensional architectures can host fragile or short-range spin liquids and may furnish new routes toward deconfined quantum liquids upon quantization \cite{Fancelli-2025}. The broader opportunity, then, is to move from a platform-based search to a geometry-based one: not asking only which known lattices are frustrated, but which three-dimensional simplex networks generate robust local constraints, weak harmonic selection, and symmetry-allowed internal kinetics.

An especially intriguing longer-term direction involves constraint manifolds whose coarse-grained description is not a rank-1 vector gauge field but a higher-rank tensor field. Higher-rank $U(1)$ gauge theories with generalized Gauss laws and associated multipole conservation were introduced in the field-theoretic context by Pretko, who showed that symmetric rank-2 tensor gauge fields can naturally give rise to subdimensional particle dynamics and fracton-like excitations \cite{Pretko-2017a,Pretko-2017b}. Closely related lattice realizations of fracton topological order based on subsystem symmetries and generalized Gauss laws were constructed by Vijay, Haah, and Fu, who demonstrated how restricted-mobility excitations arise in exactly solvable spin models \cite{Vijay-2016}. In such systems, the local constraints resemble generalized Gauss laws for symmetric tensors, giving rise to excitations with restricted mobility. For example, a symmetric rank-2 $U(1)$ gauge theory obeys a generalized Gauss law of the form
\begin{equation}
\partial_i \partial_j E_{ij} = \rho ,
\end{equation}
where $E_{ij}$ is a symmetric tensor electric field. This constraint enforces conservation not only of charge but also of dipole moment, severely restricting the mobility of isolated charges and leading to fracton-like behavior.

While fully stabilized fractonic spin liquids remain challenging in realistic magnetic systems, the conceptual pathway is continuous: start from a classical manifold defined by multi-site constraints, ensure that dominant quantum dynamics preserve those constraints, and exploit three-dimensional geometry to prevent immediate confinement. For a comprehensive review of fracton phases, subsystem symmetries, and their relation to generalized gauge theories, see Ref.~\cite{Nandkishore-2019}. Rank-2 $U(1)$ liquids may thus be viewed, at the level of constraint logic, as higher-moment generalizations of the Coulomb phase rather than as exotic departures from it. Realistic stabilization of such higher-rank liquids likely requires symmetry protection and carefully engineered constraint structures; generic perturbations tend to reduce higher-rank constraints to conventional rank-1 forms, rendering these phases comparatively fragile \cite{Gresista-2025,Gresista-2026}.

\subsection{Open Theoretical Problems}

Several theoretical challenges remain sharply posed.

A complementary modern lens is provided by \emph{generalized (higher-form) symmetries}, which recast gauge theories in terms of symmetries acting on extended operators and characterizes phases by generalized symmetry breaking patterns. In this language, deconfinement and confinement can be viewed through the fate of emergent 1-form symmetries (and their anomalies), offering a unifying perspective that parallels an extended Landau paradigm beyond ordinary 0-form symmetry breaking~\cite{Gaiotto-2015,Gomes-2023,Greevy-2023,Lake-2018}. While we do not rely on this framework here, it provides a useful conceptual bridge between gauge structure, topological order, and modern classification tools.

First, confinement transitions in three dimensions are incompletely understood outside special limits.
While compact U(1) gauge theories admit stable Coulomb phases, the structure of Higgs transitions, symmetry-enriched confinement, and monopole proliferation in anisotropic and multipolar settings remains active territory.

Second, the boundary between constraint-generated gauge liquids and proximate ordered phases is only partially mapped.
In many candidate materials the liquid regime occupies a finite but narrow region in coupling space.
Understanding whether such ``liquid islands'' can be widened—through dimensional tuning, disorder engineering, or coupling hierarchies—remains an open materials-theory problem.

Third, disorder itself may not merely obscure spin liquids but help stabilize them.
Random transverse fields or bond disorder can suppress incipient symmetry breaking while leaving constraint manifolds intact, potentially enlarging deconfined regimes in three dimensions.
Distinguishing disorder-stabilized liquids from disorder-dominated paramagnets is therefore both a conceptual and experimental priority.

A further open direction concerns genuinely three-dimensional chiral quantum spin liquids. While two-dimensional chiral spin liquids are conceptually well established, their three-dimensional counterparts remain much less understood: the infrared field theory is still comparatively underdeveloped, the distinction between fully gapped and algebraic chiral phases remains unsettled, and concrete microscopic realizations are scarce. Recent classification and variational studies on the pyrochlore lattice~\cite{Liu-2026} nevertheless suggest that extended pyrochlore Hamiltonians with further-neighbor exchange, ring-exchange processes, spin-anisotropic couplings, or explicit chiral interactions provide a natural microscopic arena in which such phases may arise~\cite{Lapa-2012,Gomez-2024,Kranitz-2026,Iqbal-2019,Iqbal-2017}. In particular, the quantum disordering of noncoplanar pyrochlore orders with finite scalar spin chirality appears to be a promising route to three-dimensional chiral spin liquids, and equal-time correlation diagnostics may help distinguish gauge-dominated chiral regimes from states dominated by short-range background correlations~\cite{Liu-2026}.

Finally, higher-rank gauge structures and fracton-adjacent constraints challenge the assumption that three-dimensional spin liquids are exhausted by rank-1 Coulomb phases.
If local constraints enforce conservation of dipole or higher moments rather than charge alone, the resulting tensor gauge theories could support new classes of quantum liquids whose stability relies even more strongly on dimensionality.
Whether realistic magnetic Hamiltonians can naturally generate such constraints without fine tuning is a question that bridges field theory, lattice geometry, and materials chemistry.

The conclusion is that three dimensions do not simply suppress quantum disorder; instead, they expand the algebra of possible constraints and stabilize the gauge structures they generate. If two dimensions highlighted how frustration can melt order, three dimensions illustrate how geometry and symmetry can stabilize liquid phases at finite energy scales through constraint-driven gauge structure.

A three-dimensional quantum spin liquid is not stabilized by any single ingredient; it requires a sequence of mutually reinforcing conditions:

\begin{enumerate}

\item \textbf{Local constraint enforcement}. The leading interactions must first generate a robust classical manifold, typically encoded as a simplex, ice-rule, or divergence-free condition.

\item \textbf{Constraint-preserving quantum dynamics}. The dominant transverse processes must connect configurations within this manifold, rather than drive the system out of it or induce premature symmetry breaking.

\item \textbf{Weak harmonic degeneracy lifting}. If fluctuation spectra vary only weakly across the manifold, harmonic order-by-disorder produces at most subleading selection energies.

\item \textbf{Emergent gauge restructuring}. Upon coarse graining, the constrained manifold may restructure into a compact gauge description governing the low-energy sector.

\item \textbf{Three-dimensional deconfinement}. For compact U(1) gauge structure, three spatial dimensions allow a stable Coulomb phase over a finite coupling window, in contrast to the generic confinement of the two-dimensional case.

\item \textbf{Microscopic representation structure and tunability}. The symmetry character of the local moments, the form of the allowed matrix elements, and the proximity to competing phases together determine how accessible and how broad this deconfined window is in real materials.

\end{enumerate}

\ack{
We thank Ganapathy Baskaran, Mat\'ias Gonzalez, Ciarán Hickey, Harald O. Jeschke, Itamar Kimchi, Péter Kránitz, Roderich Moessner, Giuseppe Mussardo, Karlo Penc, Jeffrey Rau, Johannes Reuther, and Subir Sachdev for helpful discussions.}

\funding{R.T. acknowledges financial support by the Deutsche Forschungsgemeinschaft through ProjectID 258499086 – SFB 1170, through the Würzburg- Dresden Cluster of Excellence on Complexity and Topology in Quantum Matter – ctd.qmat Project-ID 390858490 – EXC 2147, and through the Research Unit QUAST, Project-ID 449872909 – FOR5249. R.T. thanks IIT Madras for a Visiting Faculty Fellow position under the IoE program during which this work was initiated. The work Y.I. was performed in part at the Aspen Center for Physics, which is supported by a grant from the Simons Foundation (1161654, Troyer). This research was also supported in part by grant NSF PHY-2309135 to the Kavli Institute for Theoretical Physics and by the International Centre for Theoretical Sciences (ICTS) for participating in the Discussion Meeting - Fractionalized Quantum Matter (code: ICTS/DMFQM2025/07). Y.I. acknowledges support from the Abdus Salam International Centre for Theoretical Physics through the Associates Programme, from the Simons Foundation through Grant No.~284558FY19, from IIT Madras through the Institute of Eminence program for establishing QuCenDiEM (Project No. SP22231244CPETWOQCDHOC).}

\bibliographystyle{iopart-num}
\bibliography{bibliography}

\end{document}